\documentclass[a4paper,fleqn,usenatbib]{mnras}

\usepackage{newtxtext,newtxmath}

\usepackage[T1]{fontenc}
\usepackage{ae,aecompl}

\usepackage{graphicx}	
\usepackage{amsmath}	

\title[Century-long Light Curves of RCB Stars ]{Light Curves For Ten R Coronae Borealis Stars For Longer Than a Century: Secular Evolution, Dip Statistics, and a General Model for the Shape of Isolated Light Curve Dips}

\author[B. E. Schaefer]{
Bradley E. Schaefer$^{1}$\thanks{E-mail: schaefer@lsu.edu},
\\
$^{1}$Department of Physics and Astronomy, Louisiana State University, Baton Rouge, Louisiana, 70803, USA\\
}

\pubyear{2023}

\begin{document}
\label{firstpage}
\pagerange{\pageref{firstpage}--\pageref{lastpage}}
\maketitle

\begin{abstract}

R Coronae Borealis stars (RCBs) are cool supergiants that display non-periodic deep dips in brightness.  Recently, a group of `Hot RCB stars` has been discovered to be fast evolving across the HR diagram, as these stars leave the RCB region, with brightness changes at the rate of $\sim$1 mag/century.  Perhaps cool RCB stars can also be seen evolving, either increasing in temperature as they evolve to become Hot RCB stars, or perhaps increasing in luminosity as the stars arrive at the RCB region.  To seek these changes, the only possible method is to extract archival data going back more than a century, looking for the brightness changes associated with the evolution.  I have measured and extracted 323,464 magnitudes (mostly from the Harvard plates and from the AAVSO) for ten cool RCB stars, all with over a century for the light curves, all consistently calibrated to a modern magnitude system.  For times away from any dips, these light curves are flat to within the typical uncertainty of $\pm$0.10 mag/century.  That is, I see no significant evolution.  I also have collected a large database of light curve dips and their properties.  From this, the light curves for all the well-observed isolated dips have the same shape, featuring a flat slope for the few days immediately after the minima.  Further, I derive a general model for the shape of the light curve for all isolated RCB dips, with a simple equation accurately describing the observed recovery to maximum light.
 
\end{abstract}



\section{INTRODUCTION}

The R Coronae Borealis stars (RCBs) are cool, hydrogen-deficient, carbon-rich supergiants that are normally near-constant in brightness, yet suffer occasional randomly-timed deep minima of durations months-to-decades (Clayton 1996; 2012).  The deep minima are caused when the star ejects a thick carbon dust cloud, which temporarily obscures the photosphere of the star until the cloud slowly dissipates.  RCBs are rare, with 147 RCBs currently known in the Galaxy and the Magellanic Clouds (Tisserand et al. 2020).  

RCBs have two proposed evolutionary paths (Iben et al. 1996; Saio \& Jeffery 2002)..  The first path is with the coalescence of a helium and a carbon-oxygen white dwarf in a binary system, to form an RCB star.  The second path is when a post-AGB star undergoes a final helium shell flash and puffs up to form an RCB star.  The observational evidence heavily favours the WD-merger scenario (Clayton et al. 2007, 2011; Clayton 2012).

Most RCB stars have T$_{\rm eff}$= 5000--8000 K, but there are five RCB stars that are significantly hotter (15,000--25,000 K)  (De Marco et al. 2002).  Bateson (1978) first noticed that one of these stars, DY Cen, seemed to be undergoing a longterm secular decline in brightness. De Marco et al. (2002) found that three of these stars, DY Cen, MV Sgr, and V348 Sgr, have been fading over the last century, implying an evolutionary movement from right-to-left across the top of the HR diagram.  I have confirmed and extended their discovery (Schaefer 2016) by collecting very long light curves for four Hot RCBs from archival plates at Harvard College Observatory plus modern CCD photometry reported through the {\it American Association of Variable Star Observers} (AAVSO).  For example, DY Cen increased in temperature from 5800 K in 1906, to 7500 K in 1932 (at which time the dips in brightness ceased), to 19,400 K in 1987, and then to 24,800 K in 2010.  So we are seeing rapid steady evolutionary changes in real-time, as one of the very few such cases observed in all astronomy.

The Hot RCB stars are evolving quickly away from the cool  RCB region in the HR diagram (supergiants with $\approx$ 5000--8000 K).  This raises the question of whether we can watch a cool RCB star evolving into a Hot RCB star?  Or similarly, perhaps we can catch the evolution of the progenitor as it approaches the RCB state (Lauer et al. 2019)?  These various arrivals and departures from the RCB state have timescales of centuries and millennia, so they are potentially observable with a century-long light curve.  The only way to get information on timescales approaching a century is to use the various collections of archival photographic plates, for which the oldest and largest collection is at Harvard College Observatory, stretching back to 1889.  The RCB brightness will change due to the evolution of the luminosity, as well to the evolution of the surface temperature.  The speed of evolution might perhaps be fastest for the most massive and/or the hottest of the cool RCB stars (Lauer et al. 2019).

So a reasonable plan to seek the recent evolutionary changes of the cool RCB stars is to look for secular changes in their century-long light curves.  The Harvard plates provide good coverage over the entire sky to 16th mag and deeper from roughly 1889--1954 and 1969--1989.  To get the light curve extended to 2023, I have used the CCD data recorded by the many observers reporting to the AAVSO.  Care must be taken to cast all of the magnitudes onto a consistent scale, and to evaluate the dip frequency correctly in spite of sparse data.  This paper reports my results on secular changes in the brightness and dip-frequency for ten cool RCB stars with light curves longer than a century using Harvard and AAVSO data.  I also make various statistical studies of the dips themselves.  Further, I derive a general model of the {\it shape} of isolated dip light curves, and test this with my long-term light curves.

\section{Observations}

Ten cool RCB stars, see Table 1, have been selected based on several criteria.  For example, R CrB itself was selected for the availability of a high quality and well-observed light curve going back to 1843.  W Men, HV 12842, and SP77 65-2 were chosen for being amongst the brightest RCB stars in the LMC, so that they must be amongst the most luminous RCB stars.  And I chose V3795 Sgr, VZ Sgr, and IRAS 05374-6355 as currently being the hottest of the cool RCBs, so perhaps most likely to be starting to evolve more quickly.  I selected UX Ant, V742 Lyr, and Z UMi simply for being relatively bright and inside the sky region for which data are publicly available.  I selected against Galactic RCB stars that were relatively faint or in crowded fields.  My target stars are listed in Table 1 in the order as from the {\it General Catalog of Variable Stars}, and then alphabetically by designation for those with no GCVS name.  

\begin{table*}
	\centering
	\caption{Target RCB stars and Journal of Observations}
	\begin{tabular}{llrrrrrlrrr} 
		\hline
		RCB star & Galaxy  &  $\mu$ & $B_{\rm 1950}$ & $M_{\rm B}$ & $T_{\rm eff}$ &  $\log [\epsilon(H)]$     &    Years    &    B mags    &    Visual mags    &    Max mags  \\
		\hline
UX Ant	&	Milky Way	&	$\sim$17	&	12.52	&	$\sim$~-4.8	&	7000	&	6.6	&	1890	 --	2023	&	1342	&	1979$^a$	&	1064	\\
R CrB	&	Milky Way	&	10.7	&	6.75	&	-4.1	&	6900	&	6.6	&	1784	 --	2023	&	...	&	298868	&	24696$^b$	\\
V742 Lyr	&	Milky Way	&	15.8	&	13.17	&	-2.9	&	$\lesssim$5000	&	...	&	1893	 --	2022	&	935	&	...	&	562	\\
W Men	&	LMC	&	18.5	&	14.25	&	-5.4	&	6700	&	6.0	&	1890	 --	2023	&	1256	&	2377$^a$	&	959	\\
VZ Sgr	&	Milky Way	&	14.8	&	11.08	&	-4.8	&	7000	&	6.2	&	1889	 --	2023	&	1320	&	7308$^a$	&	820	\\
V3795 Sgr	&	Milky Way	&	$\sim$15	&	12.07	&	$\sim$~-5.9	&	7000	&	$<$4.1	&	1889	 --	2023	&	699	&	4195$^a$	&	508	\\
Z UMi	&	Milky Way	&	13.3	&	12.01	&	-1.7	&	5250	&	$<$7	&	1892	 --	2023	&	2738	&	...	&	1444	\\
HV 12842	&	LMC	&	18.5	&	14.12	&	-4.6	&	...	&	6.3	&	1900	 --	2011	&	131	&	...	&	59	\\
IRAS 05374-6355	&	LMC	&	18.5	&	...	&	...	&	...	&	...	&	1900	 --	2011	&	113	&	...	&	0	\\
SP77 65-2	&	LMC	&	18.5	&	15.44	&	-3.4	&	...	&	...	&	1894	 --	2011	&	297	&	...	&	223	\\
		\hline
	\end{tabular}
	
$^a$These visual magnitudes were used only to define the times of dips.
$^b$The number is for the number of nights that are certainly at maximum light, that is, this light curve has been nightly averaged.

\end{table*}

Various fundamental properties of the target stars are collected into Table 1.  The second column of Table 1 marks the host galaxy, while the third column gives the distance modulus ($\mu$) in magnitudes.  The distances are from {\it Gaia} DR3 (Gaia Collaboration et al. 2023), Montiel et al. (2018), and Kijbunchoo et al. (2011), while the LMC RCBs have the distance modulus from Freedman et al. (2001).  The next column give the average $B$ magnitude for the year 1950, as derived later in this paper.  The fifth column gives the calculated absolute magnitude in $B$, with the addition of extinction values taken from Schlafly \& Finkbeiner (2011).  The next column lists the effective surface temperature, in degrees K, as taken from Jurcsik (1996) and Kijbunchoo et al. (2011).  The seventh column lists the hydrogen abundance in logarithmic units, $\log [\epsilon(H)]$, as calculated by Pandey, Hema, \& Reddy (2021).  Measured values were collected from Pandey et al. (2021), Kipper \& Klochkova (2006), and Jurcsik (1996).

Nearly all of my collected observations were taken from either of two sources; the Harvard archival photographic plates and the AAVSO observers.
The first major source is magnitudes recently measured from the archival photographic plates at the Harvard College Observatory.  This archive contains the majority of the world's deep plates with wide coverage.  Indeed, the coverage is the entire sky, north and south, recording all stars to 14 mag with thousands of plates, and recording all stars to 16 mag with hundreds of plates, and recording many stars to 18 mag with tens of plates.  Harvard has intense coverage from 1889 to 1954 and from around 1969 to 1989.  Almost all of these plates return magnitudes in the modern B magnitude system, while the few `red' and `yellow' plates are ignored here.  The Harvard plates are an awesome and unique source, providing thousands of magnitudes for all my RCB stars with a photometric accuracy of typically 0.10 mag.  Importantly for my project, for all the stars other than R CrB itself, Harvard plates provide the {\it only} means in the world for getting a light curve from 1889 into the 1930s.

The magnitudes from the Harvard plates have been derived with two methods.  The first method is the traditional direct by-eye visual comparison of stellar image sizes as made by looking at the plate on a light table through a magnifying loupe.  The observer makes comparisons of the RCB star image radius versus the radii of nearby comparison stars of known magnitude.  This method returns a photometric accuracy of typically 0.10 mag under ordinary good conditions.  Full details of the measure of the Harvard plates are given in Schaefer (2016).  The second method is produced by the huge program called {\it Digital Access to a Sky Century at Harvard} (DASCH) run by the Harvard College Observatory (Grindlay et al. 2012), J. Grindlay Principal Investigator.  The DASCH program has been making high-quality digital scans of all Harvard plates, with roughly half completed to date.  These scans are available publicly on-line.  Further, the DASCH program has analyzed all point sources on all plates with a very sophisticated photometry analysis (Tang et al. 2013) so as to create a light curve for every star.  These light curves are publicly available on-line{\footnote{http://dasch.rc.fas.harvard.edu/lightcurve.php}}.  The typical photometric accuracy for the DASCH magnitudes is 0.10 mag for good conditions.  Importantly, both by-eye and DASCH magnitudes have been measured many times to have a close-to-zero offset from each other (and from independent photometry) and with effectively identical one-sigma measurement errors.  Importantly, both by-eye and DASCH magnitudes are made with APASS comparison stars in the Johnson B-magnitude system, and the plate material has spectral sensitivity effectively identical to the Johnson B-magnitude system, so the resultant magnitudes are all confidently in the modern B-magnitude system.  These two points mean that both by-eye and DASCH photometry can be seamlessly combined with modern CCD magnitudes to form one consistent very-long-term light curve.

For completeness, the DASCH light curves include magnitudes that should not be used when top-quality photometry is required.  For example, a small fraction of the plates have obvious plate defects of many types at the position of the RCB star, and these constitute most of the outliers from a smooth light curve.  Further, I have rejected multiple-exposure plates and plates that are not the usual blue sensistive.  These non-useable magnitudes are readily recognized from the quality flags and parameters in the DASCH database.  Further, I have systematically rejected all magnitudes that are within 0.30 mag of the plate limit, and all magnitudes with a quoted one-sigma error bar of $>$0.30 mag.  I have examined by-eye all plates (looking at either the glass plate or the DASCH scan) with outlier magnitudes, and about half of the remainder.  So all my DASCH magnitudes have been put through a high quality control.  This procedure is `painful' for R CrB itself, because all of the plates at maximum light (with $B$$\sim$6.75) are brighter than the DASCH limit of near 8.0 mag.  That is, for stars brighter than this limit, DASCH has problems with calibration stars that are usually far away where various field effects create systematic errors.  The result is that the DASCH light curve for R CrB has the only reliable magnitudes for when it is in dip with $B$$\gtrsim$9 mag.  Fortunately, for the very long light curve of R CrB, we can use the extensive visual light curve, which is available nearly continuously from 1843 to today, plus 1784--1796.

My second major source of magnitudes is from the AAVSO, as collected from observers and telescopes worldwide since 1911.  I have taken AAVSO magnitudes from four of their primary sets of data.  The first AAVSO data set is their archived CCD magnitudes in the modern B band, mostly from the 1990s to 2023.  These are the very large number of magnitudes taken through a B-band filter, usually with relatively small telescopes, by hundreds of observers worldwide, all with the the usual standard procedures.  This data set provides the bulk of the RCB coverage from the 1990s to 2023.  The AAVSO data sets are freely available on-line{\footnote{https://www.aavso.org/data-download}}.  The second AAVSO data set is their {\it AAVSO Photometric All-Sky Survey} (APASS).  This survey covers all stars in the sky, north and south, down to roughly 17 mag with good photometric accuracy in the B and V bands.  With this, APASS provides a few recent magnitudes for my RCB stars.  APASS also provides the magnitude calibration and comparison stars used for all the other observers (the DASCH project, my by-eye magnitudes from the Harvard plates, and the AAVSO CCD observers).  The third AAVSO data set that I have used is the collection of visual magnitudes for R CrB itself as taken widely from the published literature.  This third data set is what extends the R CrB light curve from the founding of the AAVSO in 1911 back all the way to Sir John Herschel in 1843.  These visual magnitudes have identical color terms as later visual data (the human eye's spectral sensitivity does not vary) and the comparison stars are close to the modern values, so the pre-AAVSO light curve is essentially identical in nature as the normal AAVSO magnitudes reported after 1911.  These archival visual magnitudes are recorded in the same AAVSO database for R CrB.  Importantly, the typical one-sigma uncertainty for a single visual magnitude estimate of R CrB is $\pm$0.15 mag, yet when combined with many other visual measures within the same time interval, the uncertainty on the average becomes $<$0.05 mag for all of the seasonal averages.  Importantly, the intrinsic variations of the target stars is always $>$0.1 mag (due to pulsations and residual dips), so the accuracy of the time-binned visual light curve is already more than good enough for measuring any evolutionary effects.  The fourth AAVSO data set that I have used is the vast collection of visual observations for four other RCB stars.  These observations have mainly been used to specify the time intervals that the RCB star is certainly not inside a dip, for purposes of dip statistics.

I also used a set of published visual magnitudes from 1784--1796 from before the original discovery of R CrB by Pigott \& Engelfield (1797).  These observations are reported as explicit comparisons with multiple comparison stars.  To convert to modern visual magnitudes (i.e., directly comparable to the AAVSO visual magnitudes), I first converted the Johnson V magnitudes for each comparison star to visual magnitudes by the measured relation of Stanton (1999).  Then Pigott's comparisons were applied to get a visual magnitude for R CrB.  The light curve shows two intervals (1784--1785 and 1796) over which the star was apparently at maximum light.  These magnitudes are valuable because they extend my  at-maximum light curve by 59 years, which makes for a substantially more sensitive measure of any long-term secular trend.

An additional small source of magnitudes was the ASAS-3 survey\footnote{All Sky Automated Survey, \url{http://www.astrouw.edu.pl/asas/?page=aasc}} for IRAS 05374-6355 from 2002-2011.  The reason for needing these magnitudes is that I do not have any Harvard or AAVSO magnitudes after 1989.  Apparently, the star was deep in minima for most of the time.

With these sources of magnitudes, I have collected century-long light curves for the ten RCB stars.  The magnitudes and coverage are tabulated in Table 1.  The last three columns give the total number of magnitudes in the B-band (from Harvard plates and AAVSO CCD measures), in the visual band, and the total number of magnitudes for the RCB star that are certainly at maximum.

\section{RCB Light Curves}

I have collected magnitudes for each of my ten cool RCB stars, each with longer than a century of coverage.  All individual magnitudes are given in the Table 2, for which most of the data lines appear only in the Supplementary Material.  (However, the full AAVSO light curve for R CrB itself, plus the later visual magnitudes for four RCB targets used for extending the dip statistics, are publicly available in the AAVSO on-line database, and so are not included in the Table.)  The first column identifies the RCB star, and then the next two columns give the time of each observation in Julian date (JD) and fractional year.  The fourth column identifies the source of the magnitude.  For the Harvard data, this is given as the letters for the plate series followed by the plate number.  The fifth column gives the B-band magnitude and its 1-sigma error bar.  The last column gives my evaluation of whether the RCB star is at maximum light (`Max'), where I have taken a very conservative criterion.  If star cannot be confidently stated to be at maximum, then I designate the measure by `Dip', even though the star might be close to maximum light.

\begin{table}
	\centering
	\caption{RCB magnitudes (full table with 8895 magnitudes is in the Supplementary Material)}
	\begin{tabular}{llllll} 
		\hline
		RCB star & JD & Year & Source & Magnitude &  Max?   \\
		\hline
UX Ant	&	2411507.54	&	1890.38	&	B5226	&		13.37	$\pm$	0.07	&	Dip	\\
UX Ant	&	2411519.49	&	1890.41	&	B5351	&		12.80	$\pm$	0.26	&	Dip	\\
UX Ant	&	2411522.51	&	1890.42	&	B5353	&		13.50	$\pm$	0.20	&	Dip	\\
UX Ant	&	2411883.58	&	1891.41	&	B6126	&		12.30	$\pm$	0.07	&	Max	\\
UX Ant	&	2412210.68	&	1892.31	&	B7442	&		12.62	$\pm$	0.25	&	Max	\\
$\ldots$  &   &   &   &   &  \\
SP77 65-2	&	2444663.97	&	1981.16	&	DSB689	&		15.40	$\pm$	0.27	&	Max	\\
SP77 65-2	&	2444699.89	&	1981.26	&	DSB726	&		15.20	$\pm$	0.38	&	Max	\\
SP77 65-2	&	2444997.99	&	1982.08	&	DSB856	&		15.20	$\pm$	0.20	&	Max	\\
SP77 65-2	&	2447186.00	&	1988.07	&	DSB2461	&		15.00	$\pm$	0.20	&	Max	\\
SP77 65-2	&	2455563.00	&	2011.00	&	APASS	&		15.54	$\pm$	0.05	&	Max	\\
		\hline
	\end{tabular}
\end{table}

A variety of criteria can be proposed to determine when an RCB star is in a dip, or when it is at maximum light.  A primary problem is how to define maximum light, even for a perfect light curve, because the stars come out of each dip with what appears to be an asymptotic approach to some constant level that we can label as `maximum'.  (This is just due to the geometric dilution of the expanding dust cloud never going to zero extinction.)  So where should we draw the line?  This is further confused by the intrinsic pulsational variations of the parent star.  So any constant-magnitude threshold would reject either times when the star is at maximum light yet still with some substantial dust extinction (thus introducing a bias to the bright side) or times when the star is still recovering from a dip (thus providing a bias to the faint side).  The observational situation is often made greatly worse when we deal only with sparsely sampled light curves.

For identifying dips and maxima, I started off by flagging magnitudes more than 1.5 mag fainter than an approximate maximum.  This gave a list of recognized dips.  These dips occasionally consisted of multiple deep dips that had run into overlapping times.  I then expanded each time interval so as to include any measured magnitudes that were within 0.5 mag of the local maximum.  I then further expanded the time interval, perhaps by months, so that we can be highly certain that the star has recovered to maximum light.  The result is the identification of which magnitudes are certainly from maximum light (see the last column in Table 2).

The plots showing the full light curves just display the usual structure for RCB stars.  Rather, given my original purpose of looking for secular changes, I have extracted only those magnitudes that are certainly at maximum light, and these are binned them together into contiguous intervals.  For the binning intervals, I have taken only complete calendar years that have no magnitudes inside a dip.  The inside-dip identification was made extremely conservatively, with the dip intervals expanded somewhat to be sure (see last column of Table 2), and then the further selection of only pristine years provides further insolation from dips.  

I constructed binned light curves of the RCB stars certainly at maximum light, and these are the primary data product.  Each binned magnitude is a straight average of the magnitudes in each time interval.  The 1-sigma uncertainty is taken as the RMS of the included magnitudes divided by the square root of $N_{\rm obs}$, with $N_{\rm obs}$ being the number of magnitudes in the average.  In cases of large $N_{\rm obs}$, the quoted error bar is limited to 0.01 mag, in recognition that the real systematic uncertainties are something close to this.  My binned and averaged light curves at maximum are displayed in Figure 1 for R CrB and in Figure 2 for the other targets, and tabulated in Table 3.

\begin{figure}
	\includegraphics[width=\columnwidth]{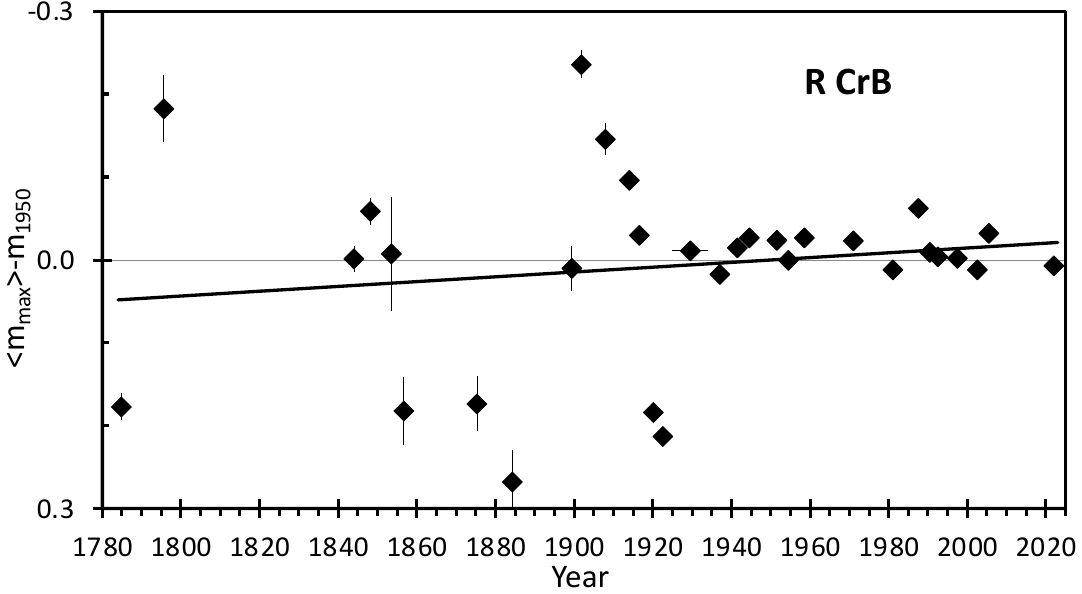}
    \caption{Binned at-maximum light curves for R CrB.  The extraction of the magnitudes at maximum produces a binned light curve that has coverage for 239 years with 298868 magnitudes, so this cannot be improved for any star even far into the future.  Before 1924, the scatter is fairly large, likely resulting from ordinary photometric problems.  After 1924, the large number of magnitudes and their good calibration makes for a scatter with an RMS of near 0.02 mag.  For my primary science question, we get a very accurate measure of any secular change due to evolution.  For the full interval of 1784--2023, the best linear fit gives a slope -0.029$\pm$0.045 mag per century, which is to say that we see no significant evolution in this best case.  This line is the thick black line, while the flat gray line provides a guide for the eye.  If we just look at the 1924--2023 century, the best-fitting line has a close-to-zero slope, indicating no significant evolution.}
\end{figure}

\begin{figure*}
	\includegraphics[width=\textwidth]{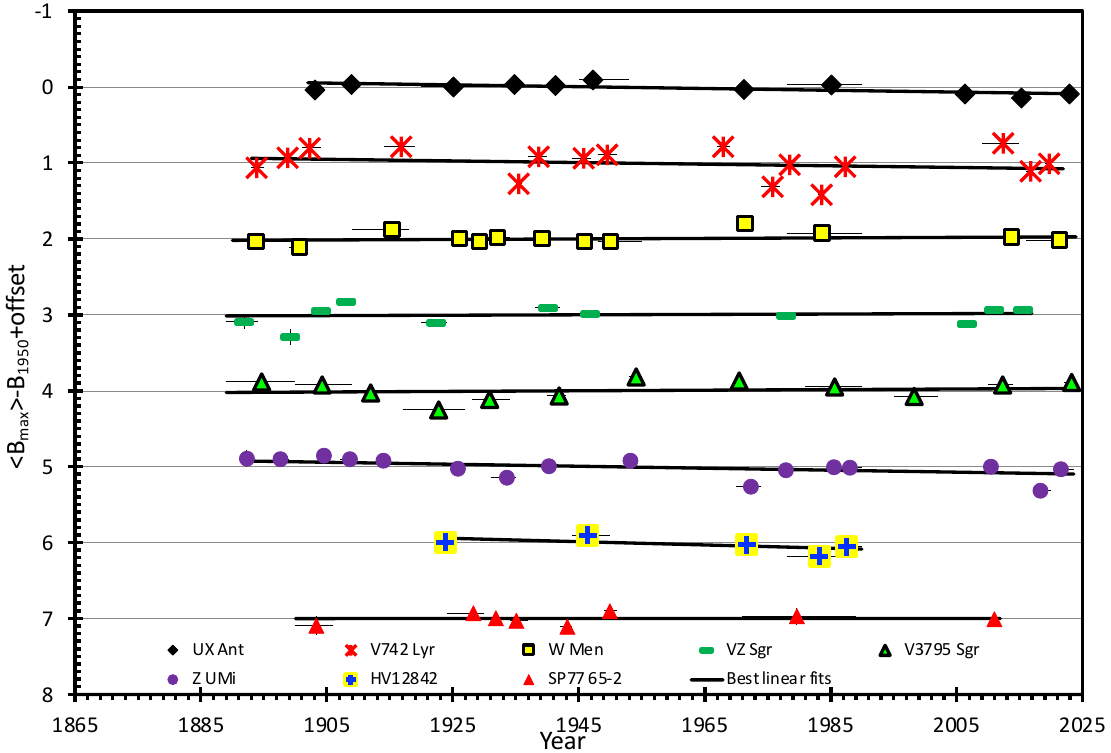}
    \caption{Binned at-maximum light curves for 8 RCB stars.  This is a plot of the light curves in Table 3 and their best line fits in Table 4.  Here are plots for eight RCB stars (each displayed with an integer offset from the $B_{\rm 1950}$ values, with the thin gray lines providing a reference for a perfectly flat line, and with the thick black lines showing the best linear fits.  Such a plot for R CrB is in Figure 1, while IRAS 05374-6355 was never seen at maximum light.  This is my primary observational result.  The idea is to use these plots and fits to seek secular brightness changes from evolution as the RCB stars either enter or leave the RCB region of the HR diagram.  The primary result is that none of the RCB stars shows any significant secular trend.  That is, these stars are all evolving at a rate such that their B magnitudes change by less than roughly 0.10 mag per century.  The individual points show variations around any best-fitting curve, far larger than any measurement or systematic errors, and I argue that these must be due to intrinsic variability of the star, for example from small unrecognized dust dips or from ordinary RCB pulsations.}
\end{figure*}

\begin{table}
	\centering
	\caption{RCB binned light curves at maximum light}
	\begin{tabular}{lllrl} 
		\hline
		RCB star & Years & $\langle Year \rangle$ & $N_{\rm obs}$ &  $B_{\rm max}$   \\
		\hline
UX Ant	&	1902--1903	&	1903.1	&	23	&	12.55	$\pm$	0.03		\\
UX Ant	&	1907--1910	&	1908.9	&	27	&	12.48	$\pm$	0.03		\\
UX Ant	&	1920--1929	&	1925.1	&	52	&	12.51	$\pm$	0.03		\\
UX Ant	&	1933--1935	&	1934.7	&	63	&	12.48	$\pm$	0.02		\\
UX Ant	&	1939--1942	&	1941.3	&	195	&	12.50	$\pm$	0.01		\\
UX Ant	&	1945--1952	&	1947.3	&	214	&	12.42	$\pm$	0.01		\\
UX Ant	&	1970--1972	&	1971.2	&	17	&	12.55	$\pm$	0.05		\\
UX Ant	&	1978--1989	&	1985.1	&	114	&	12.49	$\pm$	0.02		\\
UX Ant	&	2006	&	2006.3	&	36	&	12.61	$\pm$	0.01		\\
UX Ant	&	2015	&	2015.3	&	11	&	12.66	$\pm$	0.01		\\
UX Ant	&	2019--2023	&	2022.9	&	82	&	12.61	$\pm$	0.01		\\
R CrB	&	1784--1785	&	1784.8	&	3	&	6.24	$\pm$	0.02	$^a$	\\
R CrB	&	1795	&	1795.6	&	8	&	5.88	$\pm$	0.04	$^a$	\\
R CrB	&	1843--1844	&	1844.0	&	30	&	6.06	$\pm$	0.02	$^a$	\\
R CrB	&	1847--1849	&	1848.1	&	24	&	6.00	$\pm$	0.02	$^a$	\\
R CrB	&	1853	&	1853.5	&	9	&	6.06	$\pm$	0.07	$^a$	\\
R CrB	&	1856	&	1856.6	&	20	&	6.25	$\pm$	0.04	$^a$	\\
R CrB	&	1875	&	1875.3	&	33	&	6.24	$\pm$	0.03	$^a$	\\
R CrB	&	1882--1885	&	1884.2	&	26	&	6.33	$\pm$	0.04	$^a$	\\
R CrB	&	1898--1899	&	1899.4	&	33	&	6.07	$\pm$	0.03	$^a$	\\
R CrB	&	1901--1902	&	1901.8	&	210	&	5.83	$\pm$	0.02	$^a$	\\
R CrB	&	1906--1908	&	1907.9	&	80	&	5.92	$\pm$	0.02	$^a$	\\
R CrB	&	1913--1914	&	1914.0	&	472	&	5.97	$\pm$	0.01	$^a$	\\
R CrB	&	1916	&	1916.5	&	219	&	6.03	$\pm$	0.01	$^a$	\\
R CrB	&	1919--1920	&	1920.1	&	399	&	6.25	$\pm$	0.01	$^a$	\\
R CrB	&	1922	&	1922.5	&	312	&	6.28	$\pm$	0.01	$^a$	\\
R CrB	&	1925--1933	&	1929.6	&	2976	&	6.05	$\pm$	0.01	$^a$	\\
R CrB	&	1936--1937	&	1937.0	&	634	&	6.08	$\pm$	0.01	$^a$	\\
R CrB	&	1941	&	1941.5	&	271	&	6.05	$\pm$	0.01	$^a$	\\
R CrB	&	1944	&	1944.5	&	269	&	6.04	$\pm$	0.01	$^a$	\\
R CrB	&	1951	&	1951.5	&	291	&	6.04	$\pm$	0.01	$^a$	\\
R CrB	&	1954	&	1954.5	&	313	&	6.06	$\pm$	0.01	$^a$	\\
R CrB	&	1957--1959	&	1958.5	&	924	&	6.04	$\pm$	0.01	$^a$	\\
R CrB	&	1970--1971	&	1971.0	&	705	&	6.04	$\pm$	0.01	$^a$	\\
R CrB	&	1979--1982	&	1981.0	&	1434	&	6.07	$\pm$	0.01	$^a$	\\
R CrB	&	1987	&	1987.5	&	358	&	6.00	$\pm$	0.01	$^a$	\\
R CrB	&	1990	&	1990.5	&	362	&	6.05	$\pm$	0.01	$^a$	\\
R CrB	&	1992	&	1992.5	&	362	&	6.06	$\pm$	0.01	$^a$	\\
R CrB	&	1997	&	1997.5	&	362	&	6.06	$\pm$	0.01	$^a$	\\
R CrB	&	2002	&	2002.5	&	362	&	6.07	$\pm$	0.01	$^a$	\\
R CrB	&	2004--2006	&	2005.5	&	1082	&	6.03	$\pm$	0.01	$^a$	\\
R CrB	&	2021--2023	&	2022.0	&	695	&	6.07	$\pm$	0.01	$^a$	\\
V742 Lyr	&	1893--1894	&	1893.8	&	3	&	13.23	$\pm$	0.01		\\
V742 Lyr	&	1897--1899	&	1898.8	&	5	&	13.10	$\pm$	0.09		\\
V742 Lyr	&	1901--1903	&	1902.3	&	10	&	12.97	$\pm$	0.09		\\
V742 Lyr	&	1914--1918	&	1916.7	&	11	&	12.95	$\pm$	0.09		\\
V742 Lyr	&	1935	&	1935.5	&	68	&	13.44	$\pm$	0.03		\\
V742 Lyr	&	1937--1939	&	1938.6	&	125	&	13.09	$\pm$	0.02		\\
V742 Lyr	&	1944--1946	&	1945.8	&	47	&	13.11	$\pm$	0.06		\\
V742 Lyr	&	1948--1950	&	1949.5	&	68	&	13.06	$\pm$	0.03		\\
V742 Lyr	&	1967--1968	&	1967.9	&	4	&	12.95	$\pm$	0.03		\\
V742 Lyr	&	1974--1976	&	1975.8	&	42	&	13.48	$\pm$	0.04		\\
V742 Lyr	&	1978	&	1978.5	&	13	&	13.19	$\pm$	0.05		\\
V742 Lyr	&	1983	&	1983.5	&	23	&	13.59	$\pm$	0.04		\\
V742 Lyr	&	1985--1989	&	1987.3	&	89	&	13.22	$\pm$	0.03		\\
V742 Lyr	&	2009--2014	&	2012.4	&	27	&	12.91	$\pm$	0.02		\\
V742 Lyr	&	2016	&	2016.8	&	8	&	13.28	$\pm$	0.03		\\
V742 Lyr	&	2019	&	2019.7	&	29	&	13.18	$\pm$	0.02		\\
W Men	&	1890--1895	&	1893.8	&	6	&	14.28	$\pm$	0.10		\\
W Men	&	1899--1901	&	1900.6	&	8	&	14.35	$\pm$	0.04		\\
W Men	&	1909--1917	&	1915.3	&	8	&	14.13	$\pm$	0.05		\\
W Men	&	1924--1927	&	1926.0	&	48	&	14.24	$\pm$	0.02		\\
W Men	&	1929	&	1929.2	&	42	&	14.28	$\pm$	0.02		\\
		\hline
	\end{tabular}
\end{table}
\begin{table}
	\centering
	\contcaption{RCB binned magnitudes}
	\label{tab:continued}
	\begin{tabular}{lllrl} 
		\hline
		RCB star & Years & $\langle Year \rangle$ & $N_{\rm obs}$ &  $B_{\rm max}$   \\
		\hline

W Men	&	1930--1933	&	1932.1	&	129	&	14.23	$\pm$	0.02		\\
W Men	&	1938--1939	&	1939.2	&	108	&	14.24	$\pm$	0.01		\\
W Men	&	1945--1946	&	1945.9	&	169	&	14.28	$\pm$	0.01		\\
W Men	&	1948--1954	&	1950.0	&	140	&	14.28	$\pm$	0.02		\\
W Men	&	1970--1971	&	1971.4	&	6	&	14.04	$\pm$	0.09		\\
W Men	&	1978--1989	&	1983.6	&	48	&	14.17	$\pm$	0.04		\\
W Men	&	2013	&	2013.7	&	15	&	14.23	$\pm$	0.01		\\
W Men	&	2016--2023	&	2021.4	&	42	&	14.26	$\pm$	0.02		\\
VZ Sgr	&	1889--1893	&	1891.9	&	12	&	11.17	$\pm$	0.10		\\
VZ Sgr	&	1898--1899	&	1899.2	&	17	&	11.37	$\pm$	0.11		\\
VZ Sgr	&	1903--1904	&	1904.0	&	68	&	11.03	$\pm$	0.03		\\
VZ Sgr	&	1907--1908	&	1908.1	&	59	&	10.91	$\pm$	0.02		\\
VZ Sgr	&	1920--1923	&	1922.3	&	71	&	11.18	$\pm$	0.03		\\
VZ Sgr	&	1938--1941	&	1940.2	&	124	&	10.99	$\pm$	0.03		\\
VZ Sgr	&	1945--1948	&	1946.8	&	68	&	11.07	$\pm$	0.03		\\
VZ Sgr	&	1971--1982	&	1978.0	&	27	&	11.09	$\pm$	0.04		\\
VZ Sgr	&	2006	&	2006.6	&	9	&	11.20	$\pm$	0.02		\\
VZ Sgr	&	2010--2011	&	2011.0	&	11	&	11.01	$\pm$	0.02		\\
VZ Sgr	&	2015--2016	&	2015.6	&	11	&	11.02	$\pm$	0.05		\\
V3795 Sgr	&	1889--1899	&	1894.6	&	29	&	11.95	$\pm$	0.04		\\
V3795 Sgr	&	1900--1908	&	1904.3	&	22	&	12.00	$\pm$	0.05		\\
V3795 Sgr	&	1910--1913	&	1911.9	&	32	&	12.10	$\pm$	0.03		\\
V3795 Sgr	&	1917--1926	&	1922.7	&	83	&	12.32	$\pm$	0.03		\\
V3795 Sgr	&	1928--1933	&	1930.8	&	150	&	12.19	$\pm$	0.02		\\
V3795 Sgr	&	1940--1942	&	1941.9	&	103	&	12.14	$\pm$	0.02		\\
V3795 Sgr	&	1953--1954	&	1954.1	&	7	&	11.89	$\pm$	0.09		\\
V3795 Sgr	&	1970	&	1970.5	&	9	&	11.94	$\pm$	0.09		\\
V3795 Sgr	&	1981--1989	&	1985.7	&	26	&	12.02	$\pm$	0.04		\\
V3795 Sgr	&	1995--2001	&	1998.2	&	1627	&	12.15	$\pm$	0.01		\\
V3795 Sgr	&	2010--2013	&	2012.3	&	87	&	12.00	$\pm$	0.03		\\
V3795 Sgr	&	2022--2023	&	2023.3	&	21	&	11.96	$\pm$	0.01		\\
Z UMi	&	1892	&	1892.3	&	1	&	11.90	$\pm$	0.11		\\
Z UMi	&	1897--1898	&	1897.6	&	15	&	11.91	$\pm$	0.04		\\
Z UMi	&	1904	&	1904.5	&	15	&	11.86	$\pm$	0.04		\\
Z UMi	&	1907--1909	&	1908.7	&	33	&	11.91	$\pm$	0.04		\\
Z UMi	&	1913--1914	&	1913.9	&	50	&	11.93	$\pm$	0.03		\\
Z UMi	&	1925--1926	&	1925.8	&	44	&	12.03	$\pm$	0.05		\\
Z UMi	&	1931--1934	&	1933.5	&	143	&	12.15	$\pm$	0.02		\\
Z UMi	&	1940	&	1940.3	&	64	&	12.00	$\pm$	0.02		\\
Z UMi	&	1953	&	1953.2	&	5	&	11.93	$\pm$	0.09		\\
Z UMi	&	1970--1973	&	1972.3	&	14	&	12.27	$\pm$	0.05		\\
Z UMi	&	1977--1978	&	1977.9	&	36	&	12.05	$\pm$	0.02		\\
Z UMi	&	1985	&	1985.5	&	48	&	12.01	$\pm$	0.02		\\
Z UMi	&	1987--1989	&	1988.1	&	95	&	12.02	$\pm$	0.02		\\
Z UMi	&	2010	&	2010.4	&	198	&	12.01	$\pm$	0.01		\\
Z UMi	&	2018	&	2018.3	&	32	&	12.32	$\pm$	0.04		\\
Z UMi	&	2021--2023	&	2021.6	&	60	&	12.04	$\pm$	0.02		\\
HV12842	&	1923--1924	&	1923.8	&	3	&	14.12	$\pm$	0.02		\\
HV12843	&	1944--1949	&	1946.4	&	17	&	14.02	$\pm$	0.03		\\
HV12844	&	1971	&	1971.7	&	4	&	14.15	$\pm$	0.05		\\
HV12845	&	1978--1985	&	1983.2	&	12	&	14.30	$\pm$	0.07		\\
HV12846	&	1986--1989	&	1987.6	&	18	&	14.17	$\pm$	0.07		\\
SP77 65-2	&	1900--1905	&	1903.3	&	5	&	15.53	$\pm$	0.12		\\
SP77 65-3	&	1924--1929	&	1928.3	&	58	&	15.37	$\pm$	0.03		\\
SP77 65-4	&	1930--1933	&	1931.9	&	19	&	15.44	$\pm$	0.07		\\
SP77 65-5	&	1934--1936	&	1935.1	&	73	&	15.46	$\pm$	0.03		\\
SP77 65-6	&	1942--1943	&	1943.2	&	26	&	15.55	$\pm$	0.05		\\
SP77 65-7	&	1949--1950	&	1949.9	&	12	&	15.34	$\pm$	0.07		\\
SP77 65-8	&	1971--1988	&	1979.6	&	9	&	15.41	$\pm$	0.11		\\
SP77 65-9	&	2011	&	2011	&	1	&	15.45	$\pm$	0.05		\\
		\hline
	\end{tabular}
	\\
$^a$All R CrB magnitudes are consistently in the visual magnitude system.
	
\end{table}

The light curves are modern B-band magnitudes from the Harvard archival plates plus the AAVSO CCD B-band magnitudes.  However, there is one exception, where visual magnitudes are used.  The first exception is that all of the R CrB light curve is consistently constructed in the visual band.  I could have constructed a light curve from 1889 to 2023 in the B-band, but then there would always be worries about connecting this to the pre-1889 visual light curve, and since the primary task is looking for secular trends, I sought to eliminate such questions by using only visual data for the entire 1784-2023 interval.

\section{RCB Evolution}

My primary goal is to test for secular changes of RCB stars caused by the evolution as a star either approaches or leaves the RCB region of the HR diagram.  For this, I am looking for systematic changes across the last century from the binned at-maximum light curves in Table 3 and Figs 1 and 2.  Even a brief examination of my figures shows that there are no large or obvious secular changes on any long time-scale.  Still, I should quantify this result.

I expect that evolutionary changes will be on a sufficiently long time-scale such that the changes will be smooth across one century.  That is, I should be looking for {\it linear} variations.  This is best done as a chi-square fit for the individual light curves in Table 3.  The model is taken as $B_{\rm linear}=B_{\rm 1950} + S(Y-1950)/100$, where $B_{\rm 1950}$ is the magnitude in the year 1950, $S$ is the slope in units of magnitudes per century, and $Y$ is the year of the observation.  This model has been centred in 1950 so as to make for uncorrelated errors, and the units of magnitudes per century makes for convenient and understandable values.  My use of chi-square fitting allows for the usual calculation of uncertainties on $B_{\rm 1950}$ and $S$ as being the largest changes over which the $\chi^2$ rises from its best-fitting minimum value by 1.00 for the 1-sigma error bar.  The cause of the scatter about the best linear fits could be any or all of (1) unrecognized systematic errors, (2) small or short unrecognized dust dips included in the at-maximum light curve, and (3) ordinary pulsational variations being sampled imperfectly.

With this, I have made linear fits to my binned at-maximum light curves, with the results tabulated in Table 4.  The first column is the RCB star.  I have no fit for IRAS 05374-6355, because I have never found it at maximum light at anytime since 1900.  The next two columns give the best-fitting $B_{\rm 1950}$ and $S$, along with their 1-sigma error bars calculated with the added intrinsic variance.  The fourth column is the observed RMS scatter (in magnitudes) of the averaged magnitudes about the best-fitting line.  My best-fitting lines are plotted in Figs 1 and 2.  

\begin{table}
	\centering
	\caption{Linear fits to century-long at-maximum light curves}
	\begin{tabular}{lrrrr} 
		\hline
		RCB star & $B_{\rm 1950}$ (mag) & $S$ (mag/cen) & RMS (mag) \\
		\hline
UX Ant	&	12.52	$\pm$	0.02	&	0.12	$\pm$	0.05	&	0.05	\\
R CrB	&	6.06	$\pm$	0.02	&	-0.03	$\pm$	0.05	&	0.11	\\
V742 Lyr	&	13.17	$\pm$	0.06	&	0.10	$\pm$	0.14	&	0.20	\\
W Men	&	14.25	$\pm$	0.02	&	-0.03	$\pm$	0.05	&	0.08	\\
VZ Sgr	&	11.08	$\pm$	0.04	&	-0.02	$\pm$	0.08	&	0.14	\\
V3795 Sgr	&	12.07	$\pm$	0.04	&	-0.04	$\pm$	0.10	&	0.13	\\
Z UMi	&	12.01	$\pm$	0.03	&	0.13	$\pm$	0.07	&	0.09	\\
HV 12842	&	14.12	$\pm$	0.04	&	0.21	$\pm$	0.15	&	0.08	\\
SP77 65-2	&	15.44	$\pm$	0.03	&	0.00	$\pm$	0.10	&	0.07	\\
		\hline
	\end{tabular}
	
\end{table}

Now we can look for evolution as a slope $S$ being significantly different from zero.  A positive $S$ value indicates a {\it fading} of the star, while a negative value indicates a {\it brightening} of the star.  A star approaching the RCB region on the HR diagram will come from the bottom, so it should be brightening.  A star leaving the RCB region will be shifting to higher temperature at nearly-constant luminosity, and this will increase the B-band brightness as the bolometric correction changes.  So the expectation is that any evolutionary change should result in a {\it negative} $S$ value.

From Table 4, we only see $S$ values within 2.3-sigma of zero.  The largest deviations from zero slope are all with a positive slope.  In no case can we point to any significant, or even suggestive, secular change.  So this is my result, we have nice century-long light curves at maximum for nine RCB stars, and we see no evidence of evolution, with a typical uncertainty of 0.10 magnitudes per century.

\section{Dip Statistics}

Long-term RCB light curves can also allow the measure of properties and statistics for the dips themselves.  Little work has been published on such a plan.  The only work that I can find is a study in 1996 by J. Jurcsik (Jurcsik 1996).  He was able to collect light curves for 27 RCB stars, with a median coverage of 64 years in length, for a total of 358 dips.  From this, he only derived one dip statistic, the average time between dips ($\langle \Delta T \rangle$).  Jurcsik found that this $\langle \Delta T \rangle$ is strongly correlated with relative abundance of hydrogen in the RCB stellar atmosphere.  The relation is that a low hydrogen abundance points to a short average time between dips.  Correspondingly, a high hydrogen abundance goes along with a long time between dips.  Jurcsik's excellent work provides a characteristic clue as to the dip-formation mechanism.

\subsection{Dip Dates, Durations, and Depths}

My century-long light curves can also be used to provide a thorough data set on RCB dips and their statistics.  A dip was taken to be an interval over which at least one measured magnitude was fainter than 1.5 mag below the average maximum light, and each dip is distinct from adjacent dips by the star recovering to around maximum for some time between.  Each dip consists either of single isolated dips with just one brightness minimum, or multiple dips caused by multiple mass ejections leading to complex dips with multiple local minima in brightness.  Table 5 collects some properties on all the dips (isolated and complex) that I have recognized, 174 in total.  The first column is the RCB star, and subsequent columns give the Julian date (JD) of the dip, the duration of the dip in days, and the faintest observed B magnitude ($B_{\rm min}$).  For this, I have not listed any one dip for IRAS 05374-6355, as I only see a continuous dip from 1900 to 2011.  In general, I took the JD of the dip to be the date with the deepest observed measured magnitude, which I took to be $B_{\rm min}$.  I took the dip duration to be the interval from the first to last time for which the star was more than 1.5 mag fainter than its average maximum magnitude.

\begin{table}
	\centering
	\caption{RCB dips}
	\begin{tabular}{llrr} 
		\hline
		RCB star & JD & Duration & $B_{\rm min}$   \\
		\hline
UX Ant	&	2413246	&	<438	&	>14.02	\\
UX Ant	&	2414037	&	<413	&	>14.41	\\
UX Ant	&	2415073	&	<335	&	14.20	\\
UX Ant	&	2421349	&	<690	&	>14.45	\\
UX Ant	&	2426455	&	<105	&	>15.24	\\
UX Ant	&	2428719	&	329	&	16.50	\\
UX Ant	&	2431109	&	62	&	15.00	\\
UX Ant	&	2451881	&	431	&	>15.00	\\
UX Ant	&	2452440	&	335	&	>14.60	\\
UX Ant	&	2454298	&	1155	&	>18.80	\\
UX Ant	&	2456085	&	32	&	15.28	\\
UX Ant	&	2457800	&	>108	&	16.10	\\
R CrB	&	2377016	&	246	&	>7.8	\\
R CrB	&	2395200	&	<150	&	8.00	\\
R CrB	&	2397543	&	>80	&	10.00	\\
R CrB	&	2398240	&	>4	&	8.00	\\
R CrB	&	2398791	&	270	&	10.70	\\
R CrB	&	2399793	&	37	&	10.70	\\
R CrB	&	2400196	&	162	&	11.27	\\
R CrB	&	2400987	&	40	&	12.70	\\
R CrB	&	2402137	&	2318	&	13.30	\\
R CrB	&	2405077	&	1335	&	13.90	\\
R CrB	&	2407205	&	>80	&	9.70	\\
R CrB	&	2407760	&	>40	&	13.50	\\
R CrB	&	2410497	&	29	&	7.60	\\
R CrB	&	2411865	&	>300	&	8.80	\\
R CrB	&	2413041	&	>510	&	10.30	\\
R CrB	&	2413409	&	<40	&	9.50	\\
R CrB	&	2416202	&	<140	&	7.98	\\
R CrB	&	2416289	&	18	&	8.53	\\
R CrB	&	2416991	&	<430	&	10.00	\\
R CrB	&	2418503	&	350	&	13.80	\\
R CrB	&	2418762	&	116	&	12.73	\\
R CrB	&	2418974	&	140	&	11.80	\\
R CrB	&	2419249	&	253	&	12.10	\\
R CrB	&	2419532	&	16	&	7.85	\\
R CrB	&	2420729	&	33	&	8.70	\\
R CrB	&	2421397	&	399	&	13.70	\\
R CrB	&	2422768	&	15	&	9.40	\\
R CrB	&	2422973	&	40	&	9.79	\\
R CrB	&	2423557	&	59	&	12.43	\\
R CrB	&	2424070	&	17	&	8.93	\\
R CrB	&	2427829	&	212	&	12.01	\\
R CrB	&	2429235	&	422	&	13.57	\\
R CrB	&	2430673	&	122	&	12.50	\\
R CrB	&	2431542	&	15	&	7.85	\\
R CrB	&	2431807	&	37	&	9.50	\\
R CrB	&	2431877	&	24	&	9.70	\\
R CrB	&	2432173	&	105	&	11.60	\\
R CrB	&	2432999	&	352	&	14.10	\\
R CrB	&	2434360	&	107	&	11.00	\\
R CrB	&	2437053	&	153	&	12.57	\\
R CrB	&	2437207	&	48	&	11.80	\\
R CrB	&	2437944	&	166	&	13.20	\\
R CrB	&	2438593	&	1454	&	14.85	\\
R CrB	&	2440422	&	13	&	8.03	\\
R CrB	&	2441424	&	93	&	12.40	\\
R CrB	&	2442098	&	133	&	11.45	\\
R CrB	&	2442735	&	115	&	11.13	\\
R CrB	&	2443281	&	303	&	13.44	\\
R CrB	&	2445624	&	240	&	13.83	\\
R CrB	&	2446348	&	44	&	10.48	\\
R CrB	&	2447429	&	199	&	11.45	\\
		\hline
	\end{tabular}
\end{table}

\begin{table}
	\centering
	\contcaption{RCB dips}
	\label{tab:continued}
	\begin{tabular}{llrr} 
		\hline
		RCB star & JD & Duration & $B_{\rm min}$   \\
		\hline

R CrB	&	2447774	&	28	&	8.48	\\
R CrB	&	2449297	&	28	&	8.27	\\
R CrB	&	2450042	&	324	&	13.60	\\
R CrB	&	2451121	&	105	&	8.58	\\
R CrB	&	2451203	&	54	&	9.67	\\
R CrB	&	2451422	&	132	&	13.70	\\
R CrB	&	2451901	&	48	&	12.39	\\
R CrB	&	2452700	&	51	&	12.75	\\
R CrB	&	2454948	&	3592	&	15.10	\\
R CrB	&	2458777	&	32	&	8.51	\\
R CrB	&	2460113	&	53	&	10.15	\\
V742 Lyr	&	2415151	&	<732	&	>15.54	\\
V742 Lyr	&	2419308	&	>1300	&	>15.72	\\
V742 Lyr	&	2424801	&	>4070	&	>17.27	\\
V742 Lyr	&	2428343	&	<504	&	>15.52	\\
V742 Lyr	&	2430617	&	1105	&	>17.16	\\
V742 Lyr	&	2432347	&	<533	&	>15.92	\\
V742 Lyr	&	2433778	&	<621	&	16.10	\\
V742 Lyr	&	2437851	&	>31	&	>14.96	\\
V742 Lyr	&	2440390	&	>1570	&	>15.45	\\
V742 Lyr	&	2443396	&	30	&	>14.71	\\
V742 Lyr	&	2444100	&	1466	&	>15.08	\\
V742 Lyr	&	2445873	&	<350	&	>15.08	\\
V742 Lyr	&	2457253	&	<664	&	15.64	\\
V742 Lyr	&	2458010	&	334	&	15.97	\\
V742 Lyr	&	2458982	&	397	&	18.65	\\
W Men	&	2413878	&	109	&	>15.78	\\
W Men	&	2416378	&	479	&	16.00	\\
W Men	&	2417966	&	<1500	&	>16.90	\\
W Men	&	2422610	&	<2100	&	>16.27	\\
W Men	&	2423466	&	>30	&	15.83	\\
W Men	&	2425598	&	39	&	>16.96	\\
W Men	&	2427807	&	>140	&	16.13	\\
W Men	&	2430726	&	760	&	16.26	\\
W Men	&	2443931	&	50	&	16.00	\\
W Men	&	2445398	&	30	&	16.00	\\
W Men	&	2447202	&	334	&	16.80	\\
W Men	&	2450453	&	40	&	15.70	\\
W Men	&	2451880	&	30	&	15.50	\\
W Men	&	2454213	&	>410	&	>15.90	\\
W Men	&	2456900	&	180	&	16.80	\\
W Men	&	2457931	&	>240	&	>16.70	\\
VZ Sgr	&	2415533	&	>220	&	13.72	\\
VZ Sgr	&	2417361	&	>16	&	12.70	\\
VZ Sgr	&	2418917	&	2595	&	14.99	\\
VZ Sgr	&	2424655	&	144	&	14.01	\\
VZ Sgr	&	2426064	&	635	&	15.28	\\
VZ Sgr	&	2427305	&	>6	&	13.19	\\
VZ Sgr	&	2428444	&	<18	&	12.74	\\
VZ Sgr	&	2430823	&	122	&	13.73	\\
VZ Sgr	&	2431338	&	50	&	15.38	\\
VZ Sgr	&	2433071	&	<370	&	14.75	\\
VZ Sgr	&	2446225	&	770	&	14.80	\\
VZ Sgr	&	2446320	&	36	&	14.02	\\
VZ Sgr	&	2446863	&	4	&	12.80	\\
VZ Sgr	&	2447362	&	450	&	16.90	\\
VZ Sgr	&	2448528	&	13	&	13.30	\\
VZ Sgr	&	2449212	&	74	&	15.40	\\
VZ Sgr	&	2449301	&	26	&	13.30	\\
VZ Sgr	&	2450609	&	195	&	15.20	\\
VZ Sgr	&	2452160	&	116	&	14.30	\\
VZ Sgr	&	2454712	&	440	&	14.20	\\
VZ Sgr	&	2456157	&	89	&	13.90	\\
		\hline
	\end{tabular}
\end{table}

\begin{table}
	\centering
	\contcaption{RCB dips}
	\label{tab:continued}
	\begin{tabular}{llrr} 
		\hline
		RCB star & JD & Duration & $B_{\rm min}$   \\
		\hline

VZ Sgr	&	2456384	&	283	&	14.91	\\
V3795 Sgr	&	2418457	&	<640	&	13.54	\\
V3795 Sgr	&	2428395	&	94	&	13.83	\\
V3795 Sgr	&	2431619	&	385	&	13.82	\\
V3795 Sgr	&	2441153	&	>410	&	14.40	\\
V3795 Sgr	&	2444023	&	<140	&	13.60	\\
V3795 Sgr	&	2448766	&	445	&	16.30	\\
V3795 Sgr	&	2452781	&	49	&	>15.30	\\
V3795 Sgr	&	2452933	&	146	&	15.00	\\
V3795 Sgr	&	2453183	&	99	&	14.70	\\
V3795 Sgr	&	2453479	&	173	&	>16.00	\\
V3795 Sgr	&	2457064	&	441	&	18.10	\\
Z UMi	&	2413364	&	10	&	13.63	\\
Z UMi	&	2414826	&	>5	&	13.53	\\
Z UMi	&	2415925	&	>5	&	>15.55	\\
Z UMi	&	2419092	&	<200	&	13.52	\\
Z UMi	&	2422376	&	755	&	>14.76	\\
Z UMi	&	2424945	&	<130	&	13.58	\\
Z UMi	&	2425681	&	<400	&	>14.15	\\
Z UMi	&	2428016	&	180	&	>15.19	\\
Z UMi	&	2428520	&	240	&	>15.15	\\
Z UMi	&	2428931	&	120	&	>15.72	\\
Z UMi	&	2429457	&	184	&	>14.15	\\
Z UMi	&	2430076	&	872	&	16.25	\\
Z UMi	&	2431529	&	305	&	>14.82	\\
Z UMi	&	2432021	&	230	&	>14.82	\\
Z UMi	&	2432646	&	328	&	15.60	\\
Z UMi	&	2433397	&	266	&	>15.28	\\
Z UMi	&	2434120	&	<250	&	14.53	\\
Z UMi	&	2440269	&	596	&	>15.44	\\
Z UMi	&	2442490	&	598	&	>15.58	\\
Z UMi	&	2443956	&	495	&	>15.01	\\
Z UMi	&	2445208	&	150	&	>14.65	\\
Z UMi	&	2446552	&	<50	&	>14.44	\\
Z UMi	&	2454555	&	380	&	18.53	\\
Z UMi	&	2455641	&	128	&	15.20	\\
Z UMi	&	2456002	&	567	&	14.19	\\
Z UMi	&	2457249	&	480	&	16.53	\\
Z UMi	&	2458821	&	274	&	18.78	\\
HV12842	&	2424409	&	1482	&	>16.40	\\
HV12842	&	2427061	&	1845	&	>17.50	\\
HV12842	&	2429620	&	<750	&	15.70	\\
HV12842	&	2430611	&	<400	&	16.10	\\
HV12842	&	2433979	&	420	&	>16.70	\\
SP77 65-2	&	2413878	&	<2228	&	>17.00	\\
SP77 65-2	&	2429703	&	1244	&	>17.50	\\
SP77 65-2	&	2431850	&	1481	&	17.20	\\
SP77 65-2	&	2433979	&	>400	&	>17.00	\\
		\hline
	\end{tabular}
\end{table}

The dip properties have a variety of problems, even if we have a perfect light curve with hourly sampling.  There is no unique or standard means of identifying a dip, with this being critical when two or more dips overlap.  For overlapping dips, I have no way to say whether it is one dip with a complex structure or many dips for each minimum.  And I have no good definition of the duration, with many varied definitions possible.  Further, a deep problem is that dips return to maximum only asymptotically, and there is no good criterion for knowing when the dip ends.

The dip properties have large problems because the light curves are not perfect, being often sparsely sampled and the deepest magnitudes often being limits.  Many of the RCB star dips (but not for those of R CrB itself) have the time of true minimum missed, and hence I only have a lower limit on $B_{\rm min}$ and my date of minimum might be days or months different from the real JD of faintest light.  And the duration often can only have a poor upper or lower limit.  (For example, when I have two plates taken a day apart showing a definite dip, but the closest-in-time plates are only in the preceding and following observing season, should we call the duration as 1 day, $>$1day, or $<$320 days?  All these possible statements quote a numerical value that poorly represent the real situation, and many of the durations have similar very large uncertainties.)  Further, the sparse light curves mean that I inevitably miss many entire dips, hence making for a substantial distortion in the statistics of the dip-to-dip interval distribution.  My light curves have gaps every year due to the stars' conjunction with the Sun.  Further, my light curves suffer from the notorious Menzel Gap from roughly 1954--1969, caused by the Harvard Director stopping the plate program so he could fund his own solar research.  And my fainter RCB stars have a further long gap from the end of the Harvard plates in 1989 up to the start of the AAVSO light curve.

For the collected dips, I have calculated the median dip duration in days, $\langle D \rangle$, and the median dip depth in magnitudes, $\langle \delta \rangle$.  The use of the median statistic is a reasonable way to handle the many limits on the durations and depths.  These are tabulated in Table 6.

\begin{table*}
	\centering
	\caption{Statistics on RCB dips}
	\begin{tabular}{lrrrrrrrrr} 
		\hline
		RCB star &  $\langle D \rangle$ &   $\langle \delta \rangle$   & $N_{\rm lun}$ & $F_{\rm lun}$ & $N_{\rm season}$ & $F_{\rm season}$ & $\langle \Delta T \rangle$ &   $P_{\rm lun}$   &    $\langle T_{\rm HWHM} \rangle$    \\
		\hline
UX Ant	&	329	&	3.2	&	391	&	0.05	&	90	&	0.08	&	2200	&	0.012	&	34	\\
R CrB	&	111	&	5.2	&	1877	&	0.29	&	176	&	0.52	&	800	&	0.030	&	19	\\
V742 Lyr	&	698	&	2.9	&	393	&	0.33	&	84	&	0.46	&	2500	&	0.020	&	...	\\
W Men	&	140	&	1.9	&	302	&	0.06	&	66	&	0.15	&	2000	&	0.016	&	120	\\
VZ Sgr	&	122	&	3.0	&	392	&	0.12	&	84	&	0.26	&	1300	&	0.028	&	10	\\
V3795 Sgr	&	160	&	2.6	&	243	&	0.04	&	77	&	0.07	&	3500	&	0.010	&	370	\\
Z UMi	&	250	&	3.4	&	646	&	0.13	&	97	&	0.37	&	1300	&	0.022	&	62	\\
HV 12842	&	750	&	2.3	&	93	&	0.29	&	41	&	0.34	&	2500	&	0.015	&	...	\\
IRAS 05374-6355	&	...	&	...	&	104	&	1.00	&	42	&	1.00	&	...	&	...	&	...	\\
SP77 65-2	&	1363	&	1.9	&	119	&	0.07	&	40	&	0.16	&	2000	&	0.030	&	...	\\
		\hline
	\end{tabular}
	
\end{table*}

\subsection{Dip Frequency}

A reasonable way to quantify the dip frequency is to determine the fraction of the time that the star spends inside a dip.  This is trickier than it sounds because the frequency of light curve points is highly variable and clustered in time, and we need to avoid duplications and correlations.  So I have adopted a simple assignment that a single observation shows a dip if the star is more than 1.5 mag fainter than the average maximum magnitude.  Further, I give a lunation number for every observation, then I tallied up the number of lunations for which I have data ($N_{\rm lun}$), and also the fraction of those lunations during which the star is seen to be in a dip ($F_{\rm lun}$).  For another measure of dip frequency for a longer time-scale, I have assigned each observation to a specific yearly observing season, then I have tallied the number of observing seasons with observations for each RCB star ($N_{\rm season}$), and calculated the fraction of those seasons that contain at least one observed magnitude more than 1.5 mag below maximum ($F_{\rm season}$).  These numbers and fractions are in Table 6.

The $N_{\rm lun}$ value varies from 93--1877 lunations.  With 12.37 lunations per year, the cumulative coverage of the individual stars ranges from 7.5 years (for the faint HV 12842) to 152 years (for the bright R CrB).  For R CrB, we see near 84 per cent coverage from 1843--2023, and this usually goes through the solar gaps (as R CrB is far above the ecliptic).  We see large variations in the dip fraction for lunations, ranging from 0.04 for V3795 Sgr, to 0.29 for R CrB, to 1.000 for IRAS 05374-6355.  As for dip statistics by observing season, I cover from 40 (for SP77 65-2) to 176 years (for R CrB).  The dip fractions for seasons are always larger than for lunations (by factors of up to 2.8), because the seasons with a dip often contain lunations with no dip.  The fraction of observing seasons with dips ranges from 0.07 (V3795 Sgr) to 0.52 (for R CrB) to 1.00 (for IRAS 05374-6355).

\subsection{Average Dip-to-Dip Time}

What is the average time from dip-to-dip, $\langle \Delta T \rangle$?  This is a poorly defined question, even for a perfect light curve, because overlapping dips cannot be easily distinguished, and it is unclear whether closely spaced dips should have their small dip-to-dip times included in the average.  And $\langle \Delta T \rangle$ is not easy to measure because of the ubiquity of gaps in all light curves.  For the dips listed in Table 5, I can calculate a problematic median value of $\Delta T$, as listed in Table 6.

To determine the underlying $\Delta T$ distribution with the presence of gaps and overlaps, I have simulated a full observational history from 1889--2023, complete with seasonal gaps, the Menzel gap, and other observational gaps.  For every lunation, the model decides whether a dip starts as based on a single probability, $P_{\rm lun}$, then the simulation lays out a dip of some given duration, and the availability of observations is checked.  If two dips are `seen' to overlap in some time interval, then the lunation is still counted simply as being one dip.  For each instance of randomly chosen event start times, a list of `observed' dips is tabulated, and a $\Delta T$ distribution is created.  The specific $\Delta T$ for each dip-to-dip interval is the number of lunations between the two starts multiplied by 29.5 days.  The underlying average $\Delta T$ is just 29.37/$P_{\rm lun}$.  I construct such distributions for each single instance of randomly chosen event times, and I construct many such random instances so as to see the typical variations in the model values from instance-to-instance.  This schematic model can give the $\Delta T$ distribution of dips while accounting for the effects of gaps and overlap, all within a case for the completely random occurrence of dips.

Within this schematic model, I can look at the shape of the $\Delta T$ distribution.  First, I model the case for R CrB alone, with no gaps, and a mean dip duration of 4 lunations.  I adopt $P_{\rm lun}$=0.030, which gives an underlying $\langle \Delta T \rangle$=980 days.  With this, the typical model distribution gives a median $\Delta T$ around 820 days, while the range for the central 68.3 per cent is from 300--1600 days.  There is substantial variation from instance-to-instance due to small number statistics.  The full range of $\Delta T$ is from 148 days (i.e., five lunations) to 3800 days.  These model numbers are identical (to within the instance-to-instance variations suffered by reality) to those observed for R CrB.  That is, my simple model with random timing for dust ejection events is a good mimic of R CrB's dip times.

I have used the model to calculate $P_{\rm lun}$ values for each of my sample of RCB stars (see Table 6).  For all of the stars, the model (with random dip times) finds consistency with the observed $\langle \Delta T \rangle$ as well as with the observed range of $\Delta T$ values.  I see no unexplained upper or lower cutoffs in the observed $\Delta T$ distribution, I see no evidence of a preferred inter-dip time interval, and I see consistency with random dip times for time-scales longer than several dip durations.

\subsection{Are the Dips Random in Time?}

One question is whether the RCB dips occur randomly in time?  In the past, much research has gone into addressing this question, as reviewed by Clayton (1996).  The conclusion is that the dips occur at `irregular intervals'.  This should not be surprising because a randomness of the direction of ejection for the dust would effectively mask any underlying periodicity.  Further, there is evidence that the dust ejection occurs only during a restricted phase of the RCB pulsational cycle.  Such would have the dust ejection tied to particular conditions in the outer envelope of the star that vary with the pulsation.  But this evidence is weak, contradicted by other data for the same stars, and for one RCB requires the simultaneous use of {\it three} periodicities.

The randomness-or-periodicity can be tested by looking at the distribution of dip-to-dip intervals, $\Delta T$.  A periodicity should appear with $\Delta T$ as a constant for each RCB star, or possibly as multiples of the period.  A quasi-periodicity would produce a broadened peak, possibly with further peaks at multiples of the average period.  Random dust ejections will produce a Poisson distribution of $\Delta T$.  In practice, for all my target stars, the observed $\Delta T$ distributions do not display any singular value, nor a set of integer multiples of some period, nor with any apparent broadened peaks.  That is, I am not seeing any evidence for periodicities or quasi-periodicities.  Rather, the distribution of $\Delta T$ is always consistent with a Poisson distribution.  With allowing for gaps and overlaps, my model results (see the previous Section) shows that the dust-ejection times are random in time.  So this is all just confirmation that RCB stars have random dust ejection times, at least on time-scales longer than the dip duration.

Is the dust ejection random for time-scales of less than the dip duration?  Or are the ejections clustered together, or are they periodic on the pulsation period?  I cannot test the connection to any pulsations, because ephemerides for pulsation phase are generally not known and the pulsations change in period and they come-and-go with no discernible pattern.  For short-term clustering, the problem is to recognize singular dips that overlap.  In practice, useable statistics can only be had for R CrB itself, because only this star has a high-cadence long-term high-accuracy light curve.  In the wonderful R CrB visual light curve, successive overlapping dips can be recognized by their local minima in brightness.  For example, the long and complex R CrB dip from 2007--2017 shows 11 local minima, indicating that this is composed of 11-or-more singular dips.

The R CrB light curve distinctly shows that the singular dips are strongly clustered.  For example, the 2000.0--2020.0 (7305 days) light curve has 14 local minima, of which 11 are in a central 3510 day interval.  The probability of random dip times being so clustered is of order 0.02.  For the previous interval of 1990.0--2000.0 (3652 days), there are long stretches of R CrB at maximum light, punctuated by 4 episodes, lasting a total of 595 days, which contain 15 dips.  The probability of such clustering resulting from random event times is vanishingly small.  The point is that R CrB has its singular dips clustered on time-scales of the dip duration.  That is, if R CrB has a single dust ejection event, then the probability is relatively high that it will soon have another dust ejection event, and so on, making for multiple overlapping singular dips that appear as one long and complex dip.  This short-term clustering is independent of the long-term randomness.

\subsection{Well-observed Isolated Dips and the Depth/Duration Correlation}

The durations in Table 5 have a variety of problems for purposes of looking at the shapes of dips.  The primary problem is that multiple dips get their durations concatenated together.  This could be either due to separate singular dips running together so as to appear as one long dip, or poor sampling of the light curve can make for large errors in the tabulated durations.  So if we are looking for the statistics involving the individual shapes of dips, we should come up with a pristine sample of well-observed isolated singular dips.  For this, I have looked through all my light curves, and identified 18 such dips, as tabulated in Table 7.

\begin{table}
	\centering
	\caption{Well-Observed Isolated Singular Dips}
	\begin{tabular}{lrrrr} 
		\hline
		RCB star & $JD_{\rm min}$ & $\delta$ (mags) & $T_{\rm HWHM}$ (days) & $D$ (days)  \\
		\hline
UX Ant	&	2431108	&	2.48	&	34	&	29	\\
R CrB	&	2399793	&	4.62	&	17	&	26	\\
R CrB	&	2420729	&	2.62	&	19	&	18	\\
R CrB	&	2422768	&	3.32	&	8	&	8	\\
R CrB	&	2422973	&	3.71	&	19	&	22	\\
R CrB	&	2424070	&	2.85	&	13	&	12	\\
R CrB	&	2442095	&	5.31	&	61	&	82	\\
R CrB	&	2446347	&	4.32	&	21	&	32	\\
R CrB	&	2446499	&	1.29	&	19	&	0	\\
R CrB	&	2447778	&	2.33	&	17	&	13	\\
R CrB	&	2451904	&	6.52	&	19	&	30	\\
W Men	&	2456900	&	2.30	&	120	&	70	\\
VZ Sgr	&	2446860	&	0.93	&	15	&	0	\\
VZ Sgr	&	2448534	&	1.20	&	5	&	0	\\
V3795 Sgr	&	2428360	&	1.23	&	370	&	0	\\
V3795 Sgr	&	2431800	&	1.39	&	800	&	0	\\
V3795 Sgr	&	2448798	&	2.23	&	220	&	240	\\
Z UMi	&	2455705	&	3.09	&	62	&	57	\\
		\hline
	\end{tabular}
\end{table}

For each of the isolated dips, I tabulate four statistics.  The second column of Table 7 lists the JD of the minimum.  The next column gives the depth of the minimum, in magnitudes below $B_{\rm 1950}$.  The next column gives the half-width half-maximum (HWHM) of each dip, with $T_{\rm HWHM}$ being the time interval from the observed minimum until the brightness has recovered to half way between $B_{\rm min}$ and $B_{\rm 1950}$.  Operationally, this is easy to get from the light curves.  The last column in Table 7 gives the duration, $D$, with a slightly different definition from that used in Tables 5 and 6.  Here, I take $D$ to be the time from the minimum until the brightness is within 1.5 magnitudes of the average maximum magnitude.  This changed definition is done to isolate the physics of the light curve rise after the minimum (involving the dispersion of the dust clouds) from the different physics of the light curve decline before the minimum (involving the formation of the dust). 

The median $T_{\rm HWHM}$ values in Table 7 varies substantially from star-to-star, with $\langle T_{\rm HWHM} \rangle$ tabulated in the tenth column of Table 6.  The only RCB star with many well-observed isolated dips is R CrB, with a median HWHM duration of 19 days.  The values vary from 10 days (for VZ Sgr) to 370 days (for V3795 Sgr), over a range with a factor of 37$\times$ in size.

With sparse and imperfect data, it appears that the individual dips have $T_{\rm HWHM}$ that are approximately a constant for each star, at least as compared to the whole 37$\times$ range of star-to-star variation.  The best set of HWHM durations is for R CrB, where all-but-one have an RMS scatter of 24 per cent.  The one outlier (at 61 days) has an easy explanation as possibly being a composite of multiple dips that run together so as to not produce multiple local minima in brightness.  VZ Sgr has only two measures that can be expressed as having an RMS scatter of 7 days, which is small, or expressed as 3$\times$, which is still small compared to the overall 37$\times$ range.  V3795 Sgr has three measures, all greatly longer than any others, extending over a range of 3.6$\times$, which is small compared to the overall range of 37$\times$.  With this weak evidence, it appears that RCB isolated dips have a relatively constant duration, albeit with substantial scatter.

If dips for each RCB has constant $T_{\rm HWHM}$, then the dip shape must be similar for each star.  This is easily confirmed by eye when looking at many individual dip light curves, where they all share the same fast rise, a rounded minimum, followed by a fast brightening that changes to a slow brightening, to asymptotically return to the pre-dip level.  The time-scale $T_{\rm HWHM}$ will be governed by the initial position, velocity, and acceleration for the ejected dust, and this will be primarily a function of each star's radius and luminosity (i.e., its position on the H-R diagram).  With this, the dip shape will not change from dip-to-dip.  

A second way to look at the dip shape is to seek a correlation between the dip depth and duration for isolated dips.  Only R CrB itself has useful data.  For this, there is a highly significant correlation, where $\delta$ is roughly a linear function of $D$, with only modest scatter.  The existence of such a function implies that the {\it shape} for the R CrB isolated dips is roughly a constant from dip-to-dip.  That is, for dips light curves normalized by the depth, the $\delta$ versus $D$ function requires some specific dip shape.

\subsection{Correlations}

I have measured intrinsic dip characteristics ($\langle D \rangle$, $\langle \delta \rangle$, $P_{\rm lun}$, and $\langle T_{\rm HWHM} \rangle$) for my ten RCB stars.  And Table 1 has collected and calculated $M_{\rm B}$, $T_{\rm eff}$, and $\log [\epsilon(H)]$, with these representing the position on the HR diagram and the composition.  I have sought significant correlations between these various properties.

I have found no convincing correlations.  I could point to $\langle \delta \rangle$ versus $M_{\rm B}$, with the most luminous RCB stars having low median dip-depth, but this correlation does not pass any of the usual tests of significance.

I do not recover Jurcsik's correlation between hydrogen abundance and the frequency of dips.  Unfortunately, all-but-one of my targets have middling abundances covering only a small range of $\log [\epsilon(H)]$.  The one exception is V3795 Sgr, with very low hydrogen abundance and the lowest $P_{\rm lun}$ value, as recognized by Jurcsik and supporting his correlation.  However, the close-second-lowest $P_{\rm lun}$ value is for UX Ant with a middling hydrogen abundance.  Further, the two highest $P_{\rm lun}$ values are stars (R CrB and VZ Sgr) that are of middling hydrogen abundance.  So the best that I can say is that the correlation has large scatter.

Jurcsik's correlation has the dip frequency dominated by the hydrogen content of the atmosphere.  However, this cannot be the entire determinant of the dip frequency because DY Cen changed from an ordinary RCB star with frequent dips in the year 1906, when the dips turned off as the star heated up to 7500 K in 1932, and the dips have remained off up until 2023 with DY Cen having a surface temperature of 24,800 K; see Schaefer (2016).  So the star's surface temperature must also play a dominant role in dip frequency.  To follow this up, I have sought a correlation between $T_{\rm eff}$ and $P_{\rm lun}$.  The stars with the two low $T_{\rm eff}$ values (V742 Lyr and Z UMi) have middling $P_{\rm lun}$ values.  The two stars with the highest $P_{\rm lun}$ values (R CrB and VZ Sgr) have the same temperature as the two stars with the lowest $P_{\rm lun}$ values (UX Ant and V3795 Sgr).  So there is no correlation between $T_{\rm eff}$ and $P_{\rm lun}$, at least for ordinary cool RCB stars.

I have also sought correlations with the measured properties of individual dips on each RCB star, as taken from Table 5.  For this, I do find a strong correlation between duration and depth for the RCB stars with a substantial number of useable dips.  This correlation is highly significant, but also with large scatter.  This is always in the direction that deep dips have long durations.  Empirically, this can be represented as $D \propto \exp(\delta / \langle \delta \rangle)$.  Notably, this correlation is for all dips, most of which are composite, being composed of many isolated dips superposed to form a complex shape.  The obvious interpretation is that episodes with a long series of dust ejections will result in a long duration for the complex dip, while the piling up of the dust from the many ejections will result in a deep depth.

\section{Light Curve Shape for Isolated Dips}

A perusal of close-up light curves for RCB dips shows a wide range of dip shapes.  But it also shows that the long complex dips look to be composites of multiple short dips.  Indeed, when these contributing dips can be separated out, the component singular dips appear to have the same shape as each other and the same shape as isolated dips.  The component dips all have widely varying depths, but the {\it shapes} appear to be similar in all cases.  The weak depth/duration function for R CrB points to a fairly constant dip-shape.  This can be investigated further by collecting and normalizing the isolated dip light curves from Table 7.  Importantly, the dip shape can be derived from standard theory, and then compared to the reality of the light curves.

\subsection{Light Curve Shape for Isolated Dips}

The intrinsic shape of dips should be best seen by overplotting normalized isolated dip light curves.  With this, we can see the shape of the minimum, to determine whether the dip light curve is flat immediately following the time of deepest minimum.  We can also see how variable is the light curve shape from dip-to-dip.  And we can get an averaged dip shape for comparison with theoretical models.

How should the dip light curves be normalized?  First, the dip light curve should be shifted in time so that the time of deepest observed magnitude is at zero time.  Second, the time axis should be scaled by the $\langle T_{\rm HWHM} \rangle$ for each RCB star (as tabulated in Table 6).  In practice, this scaling will have $(t-t_0)(19/T_{\rm HWHM})$ for the horizontal axis, with $t-t_0$ being the time with respect to the time of the deepest minimum.  I have chosen the factor 19, in days, so that the scaling is unity for R CrB.  Third, the magnitude should be shifted so that the maximum light is set to zero.  Fourth, the magnitude scale should be normalized so that the minimum is at 1.0 for that particular dip.  The vertical axis will be $(B-B_{\rm 1950})/\delta$, stretching from near 1.0 at the bottom to 0.0 at the top.

I have taken the eleven dips for R CrB from Table 7, scaled the rises from minimum, and overplotted these in Fig. 3.  The scaling involved is to plot $(m-6.063)/\delta$ versus $(t-t_0)(19/T_{\rm HWHM})$ in days, with $T_{\rm HWHM}$=19 days for R CrB.  We see that all eleven scaled light curves closely match each other.  The accuracy of this shape can be quantified as being around 0.05 for times of 10, 30, and 40 days in Fig. 3.  This is dominated by the expected measurement error, so the underlying accuracy of the shape model must be substantially smaller than 0.05.  This is a direct demonstration that the R CrB isolated dips all have the same shape.

\begin{figure*}
	\includegraphics[width=\textwidth]{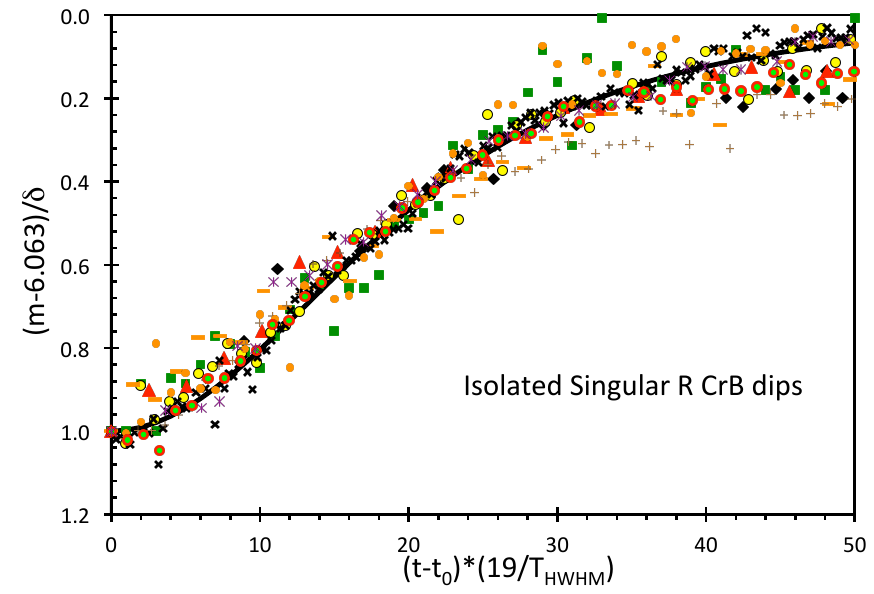}
    \caption{Overplot of eleven well-observed singular isolated dips for R CrB.  The light curves after the time of minimum are scaled in time so they all have a 19 day $T_{HWHM}$ and are scaled vertically to be for a dip with a depth of one magnitude.  Importantly, we see that all eleven dips follow the same shape.  We see that the recovery to maximum light starts out slowly (with a shallow slope in the first few days), then the recovery speeds up, with a substantially steeper slope after ten days or so.  This acceleration in the brightening is consistent with the dust cloud starting out with near-zero velocity and being immediately accelerated to high velocity by the star's radiation pressure.  Further, Equation 14 has been used to give a theoretical model for the brightening due to the expansion and dilution of the dust shell, and this model light curve is shown as the black curve.  This black curve is mostly covered by the individual data points along the centre of the light curve, and this is to say that the simple physics provides an excellent description of the `universal' dip shape.}
\end{figure*}

\begin{figure*}
	\includegraphics[width=\textwidth]{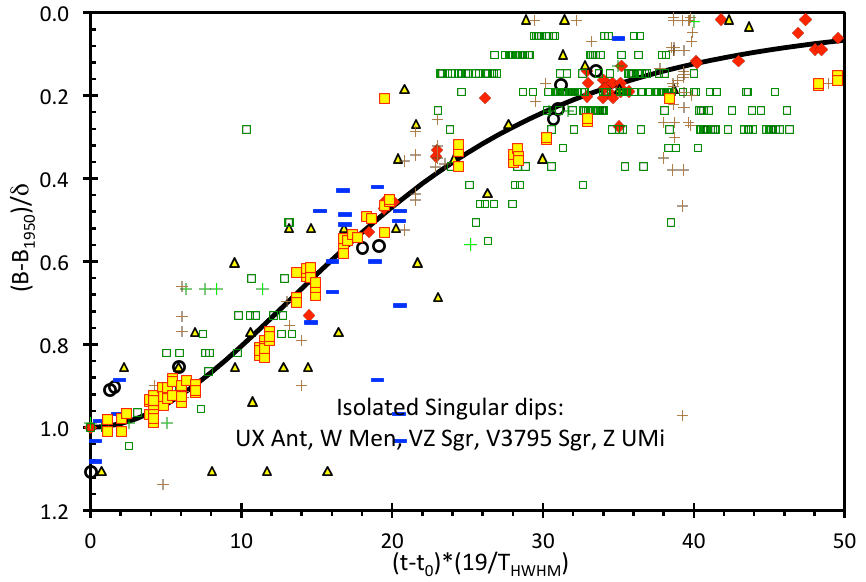}
    \caption{Overplot of eight well-observed singular isolated dips for UX Ant, W Men, VZ Sgr, V3795 Sgr, and Z UMi.  The light curves are scaled and overplotted light curves for the dips in Table 7, other than for those of R CrB itself, which appear in Fig. 3.  The construction, scales, and model curve of this plot are the same as in Figs 3 and 5, so these figures can be directly compared.  I am disappointed that these are the best data for singular isolated dips (other than those for R CrB), with the figure showing substantial scatter.  Despite the scatter, we see that the dips for the other RCB stars follows the shape of both the R CrB dips and the physics model.  While being poorer as a test than in Fig. 3, we still have a confirmation that dips from many RCB stars all apparently have the same shape.}
\end{figure*}

I have constructed a similar plot (see Fig. 4) for the eight remaining dips in Table 7.  The scatter is much larger than for R CrB, emphasizing that even the best data in the world for all other RCB stars are much poorer than for R CrB itself.  In Figure 4, I see a number of individual outlier points, and in all cases these are significantly different from many other observations of the star taken at the same time, so these are likely some form of observational error.  Part of the problem is also that half of these dips have small depth, so ordinary measurement errors get magnified on my plot.  This overall scatter looks poor, making for a relatively poor test of the constancy of the dip shapes.  Fortunately, there are many magnitudes, so the average light curve shape is accurately measured.

Importantly, the average dip shape is the same in Figs 3 and 4.  With the dip shape being the same for all the dips of all seven RCB stars, this shape appears to be of a general nature, and perhaps even `universal'.  This is the shape that should be used for general application, and the shape that needs a theory explanation.

Importantly, the light curve shape is flat for the first few days after the deepest minimum.  This will be critical for knowing the initial acceleration of the newly formed dust.


\subsection{Dip Shape from Theory}

Let us derive the theoretical shape in the light curve of the recovery from a dip.  The rapid decline of the dip is separate physics, relating to how fast the dust forms.  At some time, the dust formation is essentially complete, and that corresponds to the minimum.  After that, the ordinary ejection of the dust cloud makes for a geometric dilution of the extinction, as the dust is spread over a larger area, extending outside the surface of the star.  The primary force ($F$) acting on each dust particle is from the radiation pressure, which will have the form 
\begin{equation}
    F=kLR^{-2}, 
\end{equation}
see Equation 3 of Clayton, Geballe, \& Zhang (2013).  Here, $L$ is the star's luminosity, $R$ is the distance of the dust from the star's centre, and $k$ is some constant depending on the dust size and composition.  There will also be a force from gravity, but this will always be negligibly small, so it only acts to make the effective value of $k$ slightly smaller.  Now, the $R$ for the dust comes from the usual application of Newton's Law, so we get a non-linear differential equation for the motion;
\begin{equation}
    M_{\rm dust} \times d^2R/dt^2=kLR^{-2}.
\end{equation}
Here, $M_{\rm dust}$ is the mass of the dust particle.  The initial conditions are that the dust is at radius $R_0$ with outward velocity $V_0$ at time $t_0$.  With the solution for this equation, we can calculate the characteristic changes in the density of the expanding dust cloud, the changes in the extinction of the starlight, and thus the shape of the light curve.

The density of the dust in the cloud is $\rho_0$ at $t_0$ and $\rho$ at time $t$.  The usual geometric dilution from expansion makes for 
\begin{equation}
    \rho = \rho_0 (R/R_0)^{-2}.
\end{equation}
The optical depth of the dust cloud covering the star is
\begin{equation}
    \tau = \rho/C, 
\end{equation}
where $C$ is a constant depending of the dust size and properties, as well as the thickness of the dust shell.  In a form involving magnitudes, Bouguer's Law relates the un-extincted magnitude of the star, $m_0$, with the observed magnitude as a function of time, $m_{\rm obs}$ as
\begin{equation}
    m_{\rm obs} = m_0 +1.086 \tau .
\end{equation}
With this equation, we can calculate the brightness of the RCB star as a function of the radial position of the dust, and hence as a function of time.

Unfortunately, Equation 2 does not have any simple closed-form for its solution.  Instead, I offer three solutions.  The first solution is for the case where the acceleration is negligibly small (i.e., for $k$=0).  The second solution is for the case where the expansion is dominated by the acceleration at the early times (i.e., for $V_0$=0 and the acceleration is effectively a constant over the critical time interval).  The first and second solutions represent limiting cases that span the behavior.  The third solution is where I numerically solve Equation 2, optimizing the parameters to match the observed light curve shape from Fig. 3.

The first solution is where the acceleration is negligible.  The cloud will be moving outward with its original ejection velocity of $V_0$, and will keep coasting at this velocity.  We can define a timescale for the zero-acceleration case as
\begin{equation}
    T_0=R_0/V_0.
\end{equation}
Putting these equations together, we get a result describing the shape of the recovery after the dip minimum;
\begin{equation}
    m_{\rm obs} = m_0 + (1.086 \rho_0/C) [1+(t-t_0)/(R_0/V)]^{-2}.
\end{equation}
The radius of the dust cloud from the centre of the star at a time $t$ after the minimum time $t_0$ is 
\begin{equation}
    R = R_0 [1+(t-t_0)/T_0].
\end{equation}
With this, the {\it shape} of the recovery follows a very specific equation, which depends only on the timescale $T_0$.  The depth of the minimum, $\delta$, is $m_{\rm obs} - m_0$ for $t=t_0$, and depends only on the original dust density and properties.  We then get the general equation for single isolated dips (after the fast fading to minimum) of 
\begin{equation}
    m_{\rm obs} = m_0 + \delta [1+(t-t_0)/T_0]^{-2}.
\end{equation}
So here is a simple equation giving the universal shape for the after-minimum dip for the zero-acceleration case.  We can relate the observed timescale $T_{\rm HWHM}$ to the theoretical timescale $T_0$ by noting that $m_{\rm obs}-m_0=\delta/2$ when $t-t_0=T_{\rm HWHM}$ to get 
\begin{equation}
    T_0 = 2.41 T_{\rm HWHM}.
\end{equation}
Further, we can derive the duration-depth relation by requiring that $m_{\rm obs}-m_0=1.5$ mag when $t-t_0=D$,
\begin{equation}
    D = T_0 [(\delta/1.5)^{0.5}-1].
\end{equation}
So the limiting case with no acceleration has a simple analytic solution for the shape.

This is a very specific shape that should be universal (for the case of negligible acceleration).  The depths $\delta$ can change from dip-to-dip, and the timescale $T_0$ can change star-to-star.  Complicated dip shapes come simply from the superpositions of multiple dips, each formed by ejecta with different $t_0$ times.

This universal shape will have a variety of complications.  For example, there will be some small distortions if the dust cloud at minimum does not cover the entire star.  Further small effects will arise from the shifts of the star color as the dust clears.  And it is unclear at to when to take the time $t_0$, with small variations in the model predictions.  Further, the underlying star has pulsations with typical amplitudes of a quarter of a magnitude superposed on the dip shape.  All these effects are much smaller than the depth of the dips, with such distortions within the measurement errors.

The first solution (with $k$=0) corresponds to the scenario where the dust forms far outside the star's atmosphere, where the local gas has cooled down to the dust formation temperature.  The second solution (with $V_0$=0) corresponds to the scenario where the dust forms in the atmosphere of the star, whereupon the star's radiation rapidly accelerates the dust to high velocities.  There had long been a tension between these two scenarios, but this has been decided, with the dust formation from $V_0$=0 gas in the star's atmosphere (Clayton et al. 1992, Clayton et al. 2013).

Now let us take the other extreme limiting case, where the initial velocity is zero.  Further, let us take the acceleration to be a constant $A$, as is nearly the case where it matters only for some relatively narrow range of $R$.  We can define a timescale as
\begin{equation}
    T_A = (2R_0/A)^{0.5}.
\end{equation}
In this limit, we have 
\begin{equation}
    R/R_0 = 1+[(t-t_0)/T_A]^2.
\end{equation}
Following the derivation for the first solution, we get
\begin{equation}
    m_{\rm obs} = m_0 + \delta \{1+[(t-t_0)/T_A]^2\}^{-2}.
\end{equation}
Like for Equation 10, we can derive the duration-depth relation as
\begin{equation}
    D = T_A [(\delta/1.5)^{0.5}-1]^{0.5}.
\end{equation}
This equation adequately fits the observed depth/duration data for R CrB.  The time-scales will be related as 
\begin{equation}
    T_A = 1.55 T_{\rm HWHM}.
\end{equation}
So now we have a `universal' shape for all singular dips after the minimum.  The two free shape parameters are the depth ($\delta$) and the timescale ($T_A$).  So for the limiting case where the acceleration dominates, we have a somewhat different equation that defined the shape of the dip recovery.

The model light curve shape for the first scenario (Equation 9) and for the second scenario (Equation 14) are plotted in Fig. 5.  Critically, the two scenarios predict greatly different shapes.  The first scenario predicts that the light curve rapidly brightening immediately after the minimum, and this is definitely rejected by the data in Figs 3 and 4.  The second scenario predicts a flat light curve in the first few days after the minimum, followed by a steep brightening, and this is exactly what is seen in Figs 3 and 4.  So we have a confident empirical confirmation that the second scenario is correct, while the first scenario is rejected.

\begin{figure*}
	\includegraphics[width=\textwidth]{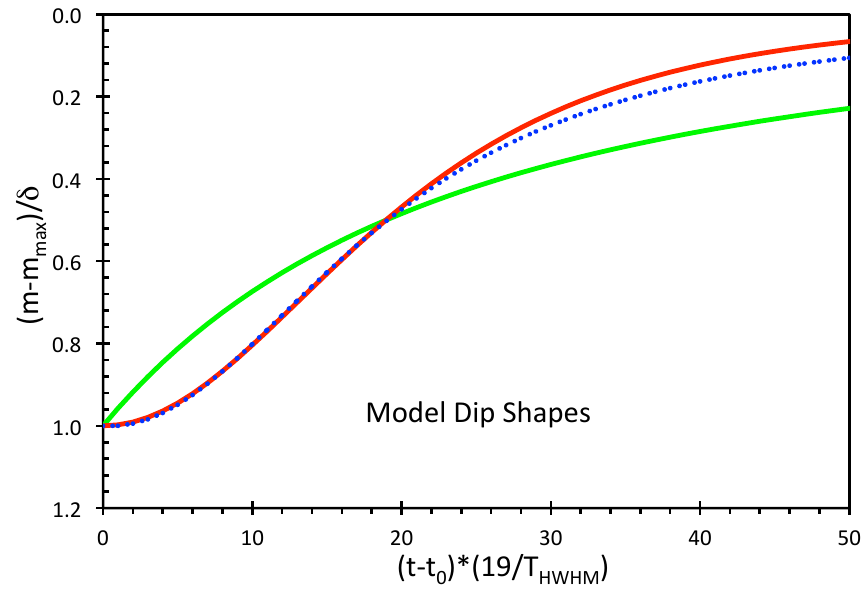}
    \caption{Model dip shapes.  The timescale for all three models have been adjusted so that $T_{HWHM}$=19 days, and so these models are directly comparable to the data in Figures 3 and 4.  The green curve (that is the lowest curve on the right side of the plot) is for the first scenario of zero acceleration (Equation 9), and has no relatively-flat initial rise.  The red curve (highest on the right side) is for the second scenario (Equation 14) where the initial velocity is zero and the acceleration does not change throughout.  This initial-low-velocity case has a relatively flat initial rise, as the dust is accelerated from low to high velocity by the star's radiation pressure.  The blue dotted curve (in the middle on the right, just below the red curve) is for the full numerical solution of Equation 2.  This case has the parameters $V_0$=0 km/s and $R_0$=75 R$_{\odot}$, as appropriate for the dust being formed in the outer stellar atmosphere, only to be accelerated away from the star by the radiation pressure.  These particular model parameters were selected so that the initial rise is fairly flat, as seen in the data (Figures 3 and 4).  The two curves for the zero-initial-velocity cases are quite close, only differing slightly in its late-time asymptotic approach to maximum.  This difference is effectively not observable due to the ordinary variations of the RCB star, for example from ordinary pulsations that can reach up to half a magnitude in amplitude.  This lack of significant difference shows that the drop of acceleration as the dust gets far away from the star has little consequence for the light curve shape.  This also means that the simple Equation 14 is an adequate description of the light curve shape.}
\end{figure*}

The first solution (with $k$=0) corresponds to the scenario where the dust forms far outside the star's atmosphere, where the local gas has cooled down to the dust formation temperature.  The second solution (with $V_0$=0) corresponds to the scenario where the dust forms in the atmosphere of the star, whereupon the star's radiation rapidly accelerates the dust to high velocities.  There had long been a tension between these two scenarios, but this has been decided, with the dust formation from $V_0$=0 gas in the star's atmosphere (Clayton et al. 1992, 2013).

Now let us take the third solution, where Equation 2 is solved numerically.  For this, we have to specify particular values within the model.  The only parameters that matter for the shape are $R_0$, $V_0$, and $A$.  With R CrB itself having a stellar radius of somewhat less than 100 R$_{\odot}$.  The initial velocity might be near-zero if the dust simply forms in the atmosphere (with $R_0$$\sim$75 R$_{\odot}$), or it might be comparable to the RCB escape velocity ($V_0$$\lesssim$100 km/s) if the dust forms in some ejected gas (perhaps around R$_{\odot}$$\sim$200 R$_{\odot}$).  The acceleration $A$ is then varied until $T_{\rm HWHM}$ matches the observations.  With these initial conditions, the numerical solution of of Equation 2 is easy, and the light curve is derived from the $R$ as a function of time with the use of Equations 3--5.  Thus, for this exact solution of Equation 2, we get a single unique function for the shape of the dips.

I have optimized the initial parameters ($R_0$, $V_0$, and $A$) to match the observed light curves in Figs 3 and 4.  Critically, the flat light curve in the first few days after minimum requires that $V_0$ must be near-zero.  In practice, the observed flat light curve immediately after the deepest minimum is predicted only for cases with $V_0$$<$10 km s$^{-1}$.  The $R_0$ and $A$ values mainly matters only with their ratio, forming $T_A$, with this being determined by the HWHM time-scale.  So the third solution is set, with $V_0$=0 and the other initial parameters set to reproduce the observed $T_{\rm HWHM}$.  This numerical solution to Equation 2 is presented as the blue dotted curve in Fig. 5. 

The primary difference amongst these three cases is that the initial rise can be linear or parabolic.  That is, the recovery to maximum can start out with some relatively fast and unchanging rate of brightening, or the recovery can start out slowly and accelerate over time to a steeper and steeper light curve.  From the physical models, the initial rise will be linear when the dust starts out with some substantial velocity (like $>$10 km/s).  The initial rise will be fairly flat when the initial velocity is low, because the dust cloud remains at nearly the same distance (and hence the same extinction) for a length of time while the acceleration from radiation pressure accumulates until the dust is moving outward fast and thinning out fast.  We see that the second and third case have very similar dip shapes.  The cases only differ in that Equation 14 makes the implicit assumption that the acceleration is a constant throughout the dip, while the numerical solution correctly has the acceleration falling as in Equation 2.  That these two solutions give a similar light curve just goes to show that the acceleration at late times is not important for the light curve.

The close similarity of the zero-initial-velocity solutions also means that, in practice, we can use the simple formula in Equation 14 to model the light curve shape after the minimum, rather than use some numerical solution that cannot be readily expressed except numerically in specific cases.  That is, Equation 14 provides an adequate physical model in a simple and convenient form.

The observations in Fig. 3 show that the rise starts fairly flat and then rapidly steepens.  This is a consequence of the substantial acceleration of the dust when it is first formed near the star.  The initial flat section of the light curve is not present if the initial velocity is much greater than 10 km/s, or if the acceleration is too low.  Indeed, we take this result as substantial evidence that the dust must start out with a very slow expansion velocity and be rapidly accelerated by the starlight's radiation pressure.  


Equation 14 works for RCB stars of all luminosity and surface temperature, and for all levels of dust formation.  The end result is a realistic physical model for the shape of the RCB dips after the initial fast decline, and the shape is expected to be universal.  The critical evidence for this model is that all the observed dip shapes for isolated dips closely fits Equation 14 (see Figs 3 and 4).

\section{Conclusions}

I have constructed century-long light curves for ten RCB stars, using 323,464 magnitudes from the Harvard plates and from the AAVSO database.  These are all calibrated into a modern magnitude system, consistent throughout the century.  I have used my century-long light curves to search for secular variations in the brightness at maximum light, as well as to get statistics on the properties and shapes of the dips.

I extracted the magnitudes for the RCB stars for times when they were certainly at maximum light, far from any dip.  These were then binned in time to get nine light curves of the stars at maximum.  These light curves show no significant secular evolution, with variations at a rate of typically $<$0.10 mag per century.

The lack of secular evolution is perhaps surprising, when seen from the point of view that some fraction of the RCB stars must be arriving or departing from the RCB region on the HR diagram over a century time.  Recall that the Hot RCB stars all show fast evolution in surface temperature resulting in brightness changes around 1 magnitude per century (Schaefer 2016).  And in particular, the now-hot RCB star DY Cen was observed to be brightening at a rate of 2.9 magnitudes per century, back at a time when $T_{\rm eff}$=5800 K.  The confident example of DY Cen demonstrates that at least some cool-R CrB stars are brightening very fast.  However, my sample includes some of the hottest of the cool-RCBs and some of the most luminous of the cool-RCBs, which are those expected to have the fastest evolution, yet they show no brightness change on a time-scale of over one century. 

I have used the light curves to extract out the dips and their measured properties.  I find only one useful correlation, and that is that the well-observed isolated dips of R CrB itself shows a highly significant relation between the dip depth and duration.

The shapes of RCB dips are often long and complex, with these apparently composed of multiple singular dips all run together.  In many of these cases, we can see that the complex dips are composites of singular dips of the same shape, with this shape being the same as for isolated dips.  So apparently, each individual dust ejection leads to a dip of the same shape, with these often superposed to appear as a long complex dip.  The general dip shape can be seen from the well-observed isolated events.  With appropriate scaling, the dip shape appears to be the same for all dips from all RCB stars with good light curves.  Critically, this shape is closely flat for the first several days after the time of minimum light.  Empirically, this shape is the same for all the isolated dips to an accuracy of roughly 5 per cent.

Theoretically, the shape of RCB dips can be derived from the acceleration of the dust and the thinning out of the dust cloud.  Critically, the predicted slope of the light curve immediately after minimum depends closely on the initial velocity of the dust on formation, where an initially-flat light curve implies an initial velocity near zero.  When combined with the empirical result that light curve is initially flat, I derive that the initial velocity is $<$10 km s$^{-1}$.  This rejects the scenario where the dust formation occurs far from the star from gas that is coasting at high velocity.  This result confirms the scenario where the dust formation occurs in zero-velocity gas that is the outer atmosphere of the RCB star.

In the end, I have an empirical and theoretical shape for isolated singular dips with the equation $m_{\rm obs} = m_0 + \delta [1+(t-t_0)/T_0]^{-2}$.  $T_0$ and $m_0$ should be a constant for a given RCB star, while $\delta$ and $t_0$ will vary from dip-to-dip.  In principle, this equation should be applicable to all RCB dips, regardless of the star's luminosity, temperature, or composition, and regardless of amount of dust formed.  And empirically, this equation applies to all well-measured isolated dips from many RCB stars.  With this, we can think that the dip shape is universal for singular dips.

\section{ACKNOWLEDGEMENTS}

I thank G. Clayton (Louisiana State University) for valuable discussions on this paper.  The {\it American Association of Variable Star Observers} ({\it AAVSO}) has provided much that was required for all of the the observers for my program, including finder charts, comparison star magnitudes (through the {\it APASS} program), programatic advice, and a data archive.  Funding for {\it APASS} has been provided by the Robert Martin Ayers Sciences Fund.  The DASCH data from the Harvard archival plates was partially supported from National Science Foundation grants AST-0407380, AST-0909073, and AST-1313370.

\section{DATA AVAILABILITY}

All data are publicly available, either in the cited literature, the cited public databases, or in the tables.


{}

\bsp	
\label{lastpage}
\end{document}